\documentclass[11pt]{article}
\pdfoutput=1
\usepackage[utf8]{inputenc}
\usepackage{caption}
\usepackage{braket,slashed,bm}
\usepackage{jheppub}
\usepackage{array,multirow}
\usepackage{amsmath}
\usepackage[normalem]{ulem}
\usepackage{calligra}
\usepackage{floatrow}
\usepackage[T1]{fontenc}
\usepackage{mathrsfs}
\usepackage{subcaption}
\usepackage{lscape}
\usepackage{hyperref}
\usepackage{graphicx}
\usepackage{booktabs}
\newfloatcommand{capbtabbox}{table}[][\FBwidth]
\def\beq{\begin{equation}}
\def\eeq{\end{equation}}
\def\bea{\begin{eqnarray}}
\def\eea{\end{eqnarray}}

\renewcommand{\to}{\rightarrow}

\def\sc{{\overline s}}

\usepackage{float}
\usepackage{enumitem}

\newcommand{\FR}{\textsc{FeynRules} }
\newcommand{\FA}{\textsc{FeynArts} }
\newcommand{\FC}{\textsc{FormCalc} }
\newcommand{\SSim}{\textsc{SMEFTSim} }
\newcommand{\MG}{\textsc{MadGraph5} }

\title{The Feynman rules for the SMEFT in the background field gauge}
\author[a]{Tyler Corbett}

\affiliation[a]{Niels Bohr International Academy,
Niels Bohr Institute, University of Copenhagen,
Blegdamsvej 17, DK-2100, Copenhagen, Denmark}

\abstract{We present a package for \FR  which derives the Feynman rules for the Standard Model Effective Field Theory up to dimension-six using the background field method for gauge fixing. The package includes operators which shift the kinetic and mass terms of the Lagrangian up to dimension-eight and including dimension-six squared effects consistently. To the best of the author's knowledge this is the first publicly available package to include dimension-six squared effects consistently. The package is validated in a partner publication by analyzing the Ward Identities at dimension-six and one-loop order. We also extend the partner work in this article by including the dimension-six squared effects further demonstrating the consistency of their implementation. In doing so we find that failure to consistently include field shifts to dimension-six squared causes a breakdown in the Ward identities implying concerns about many calculations in the literature which do not properly incorporate these effects.

The \FR files, as well as Mathematica notebooks performing the relevant calculations, can be downloaded \href{https://feynrules.irmp.ucl.ac.be/wiki/SMEFT_BGFM}{from the \FR website} and are included as ancillary files to this publication.}

\begin{document}
\maketitle

\date{\today}

\section{Introduction}
With Run 2 of the LHC finished and no signs of resonant new physics it is natural to look forward to the precision LHC program. A natural framework for this precision program is that of Effective Field Theories (EFTs). This framework allows for the  study of heavy new particles which cannot be directly, i.e. resonantly, produced at an experiment. The strength of the approach lays in its preservation of the symmetries of the assumed low energy model. The effects of new physics then manifest themselves in the form of higher dimensional operators which decouple as inverse powers of the new physics scale, $\Lambda$. This decoupling is the basis for a new power counting in inverse powers of $\Lambda$ which is in addition to power counting in gauge couplings/loops of the Standard Model (SM). The EFT most often studied for beyond the SM physics is the Standard Model Effective Field Theory, or SMEFT. In the SMEFT the higher dimensional operators of the theory are formed assuming that the Higgs boson belongs to the $SU(2)_L$ doublet of the SM. One can alternatively study the HEFT, where the Higgs boson is treated as a singlet of the full SM gauge group and the would-be goldstone bosons of the SM instead transform nonlinearly under the SM gauge group \cite{Brivio:2013pma,Brivio:2016fzo,Feruglio:1992wf,Grinstein:2007iv,Buchalla:2013rka,Contino:2010rs,Burgess:1999ha,Barbieri:2007bh,Buchalla:2012qq,Alonso:2012px}. This work focuses only on the SMEFT.

The SMEFT is formed of the SM plus a tower of operators of dimension $n\geq5$, each suppressed by the new physics scale to the power $n-4$:
\begin{equation}
\mathcal L_{\rm SMEFT}=\mathcal L_{\rm SM}+\sum_{n=5}^\infty\sum_i\frac{c_i}{\Lambda^{n-4}}\mathcal O_i\, .
\end{equation}
Where the sum over $i$ is over all operators at a given dimension. The $c_i$ are often referred to as Wilson coefficients and are the infrared parameterization of the effects of new physics at the scale $\Lambda$\footnote{In this work we will frequently absorb $\Lambda^2$ into the Wilson coefficient $c_i$ for convenience, we still make reference to the power counting in $1/\Lambda$, however.}. Neglecting flavor considerations, there is only one operator at dimension five which generates neutrino masses after electroweak symmetry breaking (EWSB). As this FeynRules package is not focused on calculations relevant to this operator we focus on dimension-six operators. Early works compiling these operators \cite{Leung:1984ni,Buchmuller:1985jz} were recently reduced to a consistent nonredundant operator basis in \cite{Grzadkowski:2010es}. This basis is often referred to as the ``Warsaw Basis'' and is the basis adopted in this paper as it is presently the most frequently used basis as well as the appeal of its making no assumptions on the normalization of the Wilson coefficients. In total the Warsaw basis of dimension-six operators is composed of 59 operators, various assumptions about flavor can result in up to 2499 free parameters (Wilson coefficients), in the flavor general case \cite{Henning:2015alf,Brivio:2017btx,Alonso:2013hga}. This work assumes the flavor general case while neglecting CP-odd operators\footnote{But not phases of nonhermitian operators}, therefore the package described in this article only contains 2493 parameters in addition to the SM parameters. These operators are outlined in Fig.~\ref{tab:op59} of Appendix~\ref{app:ops}. The general case was implemented as it allows the user to freely define their own assumptions about flavor. 

The SMEFT has become an industry for LHC and future collider studies, generating numerous global fits to the Higgs and electroweak sectors \cite{Corbett:2012dm,Corbett:2012ja,Corbett:2013pja,Corbett:2015ksa,Biekotter:2018rhp,Butter:2016cvz,Almeida:2018cld,Alves:2018nof,Ellis:2018gqa}, as well as in studies of the top quark \cite{Brivio:2019ius,Maltoni:2019aot,Degrande:2010kt,Greiner:2011tt,Buckley:2015lku,Englert:2016aei,Hartland:2019bjb,AguilarSaavedra:2010zi,Zhang:2014rja,AguilarSaavedra:2018nen}. As the LHC moves toward its precision future, serious one-loop SMEFT programs are beginning to form. A large number of studies considering only QCD loops have been done while a few also include electroweak corrections \cite{Dawson:2019clf,Hartmann:2016pil,Hartmann:2015aia,Hartmann:2015oia,Grojean:2013kd,Gauld:2015lmb,Gauld:2016kuu,Cullen:2019nnr,Cullen:2020zof}. With ever greater demand for calculations in the SMEFT framework a number of packages for simulations and calculation have been developed. \SSim \cite{Brivio:2017btx} is a \FR package which allows for tree level simulations with various flavor assumptions and the choice of two different input parameter schemes using \MG \cite{Alwall:2014hca}. The automated inclusion of one-loop QCD effects in the form of a UFO for use with \MG has also recently been made available \cite{Degrande:2020evl}. A \FR file implementing traditional $R_\xi$ gauge is also available \cite{Dedes:2017zog}. Taking the package presented here and setting all background fields to zero reproduces the $R_\xi$ gauge fixing of the SMEFT. As such this work also represents an extension of the existing $R_\xi$ gauge fixing work, in the appropriate limit, in that it includes partial $\mathcal O(1/\Lambda^4)$ effects\footnote{It should be noted, that the use of the metric to connect the gauge fixing terms in Eq.~\ref{eq:GF} means that this $R_\xi$ fixing will take a different form from that of \cite{Dedes:2017zog}.}.

Instead, this work takes the approach of the Background Field Method (BFM). Here the bosonic fields are split into background and quantum fields, $\phi\to\phi_{\rm BG}+\phi_Q$, and only the quantum fields are allowed to propagate. Since the calculations do not depend on background field propagators the background fields' need not be gauge fixed, and background gauge invariance is maintained leading to a gauge invariant effective action. This simplifies calculations and results in many nice properties, such as the preservation of the naive Ward identities, allowing for cross checks to further ensure the reliability of calculations, and allowing for Dyson summation \cite{Denner:1996gb} without breaking gauge invariance. The BFM first found use in quantum gravity \cite{DeWitt:1967ub} and in the QCD sector \cite{DeWitt:1980jv,thooft,Boulware:1980av,Abbott:1981ke}, and was subsequently introduced for the electroweak theory in \cite{Denner:1994xt}. It was then developed as a key tool for early precision calculations in the electroweak theory, for example in \cite{Denner:1996gb,Denner:1995jd,Denner:1994nn}. Gauge fixing of the SMEFT in the BFM was first introduced in \cite{Helset:2018fgq} which serves as the basis of the gauge fixing performed in this article and the associated \FR package.

This paper is organized as follows. In Section~\ref{sec:definitionslagrangian} the notation, fields, and Lagrangian are defined. Then, as a means of demonstrating the consistency and necessity of properly including dimension-six squared effects in field shifts, the two-point Ward identities of the SMEFT are shown to hold for selected dimension-six squared effects in Sec.~\ref{sec:ward}. Conclusions given in Sec.~\ref{sec:conclusions}. The appendices are used to express technical details of the implementation of the Feyman rules and their use in the Feynarts and Formcalc packages.

\section{Definitions and implementation of the Lagrangian}\label{sec:definitionslagrangian}
We follow the geometric formulation of the SMEFT presented in \cite{Helset:2020yio,Corbett:2019cwl,Hays:2020scx,Helset:2018fgq}. At this time the only dimension-eight operators implemented are of classes two, three, and four.
To outline the implementation of the SMEFT with background gauge fixing we first define necessary conventions, then define the fields and metrics, and finally the full SMEFT Lagrangian. This is also the order in which the \FR files are organized. The following is written in standard particle physics notation, but is meant to demonstrate the way in which the Lagrangian is defined from the ground up in the language of Mathematica and Feynrules.

\subsection{Conventions}
In order to implement the geometric formulation of the SMEFT Lagrangian we use the following real matrices in the place of the Pauli matrices in order to define the covariant derivative of the four-component scalar field. These are implemented in \texttt{conventions.fr}. We purposely drop the geometry inspired raised and lowered index notation of \cite{Helset:2020yio,Corbett:2019cwl,Hays:2020scx,Helset:2018fgq} for clarity, as they are not implemented in the \FR files. 
\begin{equation}
\begin{array}{ccc}
\tilde\gamma_{1IJ}=g_2\left(\begin{array}{cccc}0&0&0&-1\\0&0&-1&0\\0&1&0&0\\1&0&0&0\end{array}\right)\, ,&\ \ \ \ \ \ \ \ \ \ \ &\tilde\gamma_{2IJ}=g_2\left(\begin{array}{cccc}0&0&1&0\\0&0&0&-1\\-1&0&0&0\\0&1&0&0\end{array}\right)\, ,\\ \\
\tilde\gamma_{3IJ}=g_2\left(\begin{array}{cccc}0&-1&0&0\\1&0&0&0\\0&0&0&-1\\0&0&1&0\end{array}\right)\, ,&\ \ \ \ \ \ \ \ \ \ \ &\tilde\gamma_{4IJ}=g_1\left(\begin{array}{cccc}0&-1&0&0\\1&0&0&0\\0&0&0&1\\0&0&-1&0\end{array}\right)\, .
\end{array}
\end{equation}
with $g_1$ and $g_2$ the gauge couplings of $U(1)_Y$ and $SU(2)_L$ respectively. We also define the following for forming the metrics of \cite{Helset:2020yio,Hays:2020scx}:
\begin{equation}
\begin{array}{ccc}
\Gamma_{1IJ}=\left(\begin{array}{cccc}0&0&1&0\\0&0&0&-1\\1&0&0&0\\0&-1&0&0\end{array}\right)\, ,&\ \ \ \ \ \ \ \ \ \ \ &
\Gamma_{2IJ}=\left(\begin{array}{cccc}0&0&0&1\\0&0&1&0\\0&1&0&0\\1&0&0&0\end{array}\right)\, ,\\ \\
\Gamma_{3IJ}=\left(\begin{array}{cccc}-1&0&0&0\\0&-1&0&0\\0&0&1&0\\0&0&0&1\end{array}\right)\, ,&\ \ \ \ \ \ \ \ \ \ \ &
\Gamma_{4IJ}=-\mathbb{I}_{4\times4}\, .
\end{array}
\end{equation}
The rescaled Levi-Civita symbol is defined as,
\begin{equation}
\tilde\epsilon_{ABC}=g_2\epsilon_{ABC},\ \ \ \ \tilde\epsilon_{123}=+g_2\, ,
\end{equation}
and is defined to be zero for any $A$, $B$, $C$ equal to 4. The matrices which, in combination with the metrics defined below, rotate the scalars and vectors to mass eigenstates and canonically normalizing the kinetic terms are:
\begin{equation}\label{eq:UVrotate}
\begin{array}{ccc}
U_{AB}=\left(\begin{array}{cccc}\frac{1}{\sqrt{2}}&\frac{1}{\sqrt{2}}&0&0\\ \frac{i}{\sqrt{2}}&\frac{-i}{\sqrt{2}}&0&0\\ 0&0&c_{\bar\theta}&s_{\bar\theta}\\ 0&0&-s_{\bar\theta}&c_{\bar\theta}\end{array}\right)\, ,&\ \ \ \ \ \ \ \ \ \ \ &
V_{IJ}=\left(\begin{array}{cccc}-\frac{i}{\sqrt{2}}&\frac{i}{\sqrt{2}}&0&0\\ \frac{1}{\sqrt{2}}&\frac{1}{\sqrt{2}}&0&0\\ 0&0&-1&0\\ 0&0&0&1\end{array}\right)\, .
\end{array}
\end{equation}
The barred Weinberg angles, $c_{\bar\theta}$ and $s_{\bar\theta}$ have implicit dependence on the Wilson coefficients, which is elaborated in Appendix~\ref{app:bars}.

\subsection{Fields and Metrics}\label{sub:fieldsmetrics}
Next we define the fields and metrics in \texttt{fieldsandmetrics.fr}. The weak eigenstate fields are defined in terms of the mass eigenstates, and as such we must first define the expectation of the metrics $g_{AB}$ and $h_{IJ}$, which in combination with Eq.~\ref{eq:UVrotate} allow for the proper rotation to the mass basis. They are defined as:
\begin{eqnarray}\label{eq:expmetrics}
\langle g_{AB}\rangle&=&\delta_{AB}-4\left[c_{HW}(1-\delta_{A4})+c_{HB}\delta_{A4}\right]\frac{v_T^2}{2}\delta_{AB}-4\left[c_{HW}^{(8)}(1-\delta_{A4})+c_{HB}^{(8)}\delta_{A4}\right]\frac{v_T^4}{4}\delta_{AB}\\
&&c_{HW2}^{(8)}v_T^4\Gamma_{A44}\Gamma_{A44}(1-\delta_{A4})(1-\delta_{B4})
+\left(c_{HWB}+c_{HWB}^{(8)}\frac{v_T^2}{2}\right)\left[v_T^2\Gamma_{A44}(1-\delta_{A4})\delta_{B4}+A\leftrightarrow B\right]\, ,\nonumber\\
\langle h_{IJ}\rangle&=&\left[1+\frac{v_T^2}{4}(c_{HD}^{(8)}+c_{HD2}^{(8)})\right]\delta_{IJ}-2c_{H\Box}v_T^2\delta_{I4}\delta_{J4}+\Gamma_{AIJ}\Gamma_{A44}\frac{v_T^2}{4}(c_{HD}+v_T^2c_{HD2}^{(8)})\, .
\end{eqnarray}
Letting the $M$ stand for either metric, $g$ or $h$, the square root and inverse matrices are then defined using the perturbative definitions:
\begin{eqnarray}\label{eq:metricinvsqrts}
dM&=&M-\mathbb{I}\, \label{eq:inv1}\\
 M^{-1}&=&\mathbb{I}-dM+dM^2\, \label{eq:inv2}\\
 M^{1/2}&=&\mathbb{I}+\frac{1}{2} dM-\frac{1}{8}dM^2\, .\label{eq:inv3}
\end{eqnarray}
The function ``\texttt{WilsonLimit},'' defined in \texttt{BGFM.fr}, is then applied to ensure the series is truncated at the appropriate order set by the parameter \texttt{wilsonPower}$=$1 or 2 corresponding to $1/\Lambda^{2\times{\rm wilsonPower}}$. Note that because of these definitions this package is limited to dimension-six-squared and some dimension-eight effects and cannot be used for higher order calculations in the SMEFT expansion. The only operators included at dimension-eight at this time are those which affect the $g$ and $h$ metrics as well as the $H^8$ operator, i.e. those similar to Class 2-4 as defined in Table~\ref{app:ops}. From a theoretical standpoint it is inconsistent to use dimension-six squared effects and not include dimension-eight effects. The usage of dimension-six squared is, however, commonplace, but the Feynman rules used in these analyses do not include dimension-six squared shifts to the normalizations of fields and therefore represent incomplete calculations. This is the first public tool, to the author's knowledge, that includes these effects consistently.

The weak eigenstate fields can then be defined in terms of the physical fields. We define the four-component physical fields, $\Phi$, $\mathcal{W}_\mu$, and the ghost field, $c$ as:
\begin{equation}
\Phi=\left(\begin{array}{c}\phi^++\hat\phi^+\\ \phi^-+\hat\phi^-\\ \chi+\hat\chi\\ h+\hat h\end{array}\right)\, ,\ \ \ \ \ 
\mathcal{W}_\mu=\left(\begin{array}{c}\mathcal W_\mu^++\hat {\mathcal W}_\mu^+\\ \mathcal W_\mu^-+\hat {\mathcal W}_\mu^-\\ \mathcal Z_\mu+\hat {\mathcal Z}_\mu\\ \mathcal A_\mu+\hat {\mathcal A}_\mu\end{array}\right)\, ,\ \ \ \ \ 
c=\left(\begin{array}{c}c_{W^+}\\ c_{W^{-}}\\c_Z\\c_A\end{array}\right)\, .
\end{equation}
Here, and elsewhere unless specified, hatted fields correspond to the background fields, while unhatted correspond to quantum fields or field multiplets (e.g. $\Phi$, $\mathcal W$) - the two are distinguished by context. These are then rotated to the weak eigenstates by the definitions:
\begin{equation}\label{eq:weakeigenstatefields}
\phi_I=v_T\delta_{I4}+\langle h\rangle^{-1/2}_{IJ}V_{JK}\Phi_K\, ,\ \ \ \ \ 
W^\mu_A=\langle g\rangle^{-1/2}_{AB}U_{BC}\mathcal{W}^\mu_C\, ,\ \ \ \ \ 
u_A=\langle g\rangle^{-1/2}_{AB}U_{BC}c_C\, .
\end{equation}
Notice the true vacuum expectation value is introduced in this step as the vacuum expectation value of the fourth component of $\phi$. It is defined to be the minimum of the Higgs potential below, in Eq.~\ref{eq:LHiggs}. The standard doublet field is then defined in terms of the four component scalar as,
\begin{eqnarray}
H&=&\frac{1}{\sqrt{2}}\left(\begin{array}{c}\phi_2+i\phi_1\\ \phi_4-i\phi_3\end{array}\right)\, ,\label{eq:doublet}\\
\nonumber\\
\tilde H&=&\frac{1}{\sqrt{2}}\left(\begin{array}{c}\phi_4+i\phi_3\\ -\phi_2+i\phi_1\end{array}\right)\, .
\end{eqnarray}
With the weak eigenstate fields defined we can return to metrics and define them fully:
\begin{eqnarray}
g_{AB}&=&\delta_{AB}-4\left[c_{HW}(1-\delta_{A4})+c_{HB}\delta_{A4}\right]\frac{\phi^2}{2}\delta_{AB}-4\left[c_{HW}^{(8)}(1-\delta_{A4})+c_{HB}^{(8)}\delta_{A4}\right]\frac{\phi^4}{4}\delta_{AB}\nonumber\\
&&-c_{HW2}^{(8)}(\phi_I\Gamma^A_{IJ}\phi_J)(\phi_L\Gamma^B_{LK}\phi_K)(1-\delta_{A4})(1-\delta_{B4})\\
&&+\left(c_{HWB}+c_{HWB}^{(8)}\frac{\phi^2}{2}\right)\left[(\phi_I\Gamma^I_{IJ}\phi_J)(1-\delta_{A4})\delta_{B4}+A\leftrightarrow B\right]\, ,\nonumber\\
h_{IJ}&=&\left[1+\frac{\phi^4}{4}(c_{HD}^{(8)}+c_{HD2}^{(8)})\right]\delta_{IJ}-2c_{H\Box}\phi_I\phi_J+\frac{1}{4}(c_{HD}+\phi^2 c_{HD2}^{(8)})\Gamma^A_{IJ}\Gamma^A_{KL}\phi_K\phi_L\, .
\end{eqnarray}
Here $\phi^2\equiv\sum_I\phi_I\phi_I$ with $\phi_I$ defined as in Eq.~\ref{eq:weakeigenstatefields}, and $\phi^4=(\phi^2)^2$. Again, the inverses of these metrics are defined perturbatively with Eqs.~\ref{eq:inv1}--\ref{eq:inv3}. From here we define the covariant derivative of $\phi$ and the field stength associated with $W$, this allows us to implement the geometric formulation of the class three and four operators to $\mathcal{O}\left(\frac{1}{\Lambda^4}\right)$. They are defined as:
\begin{eqnarray}
(D^\mu\phi)_I&=&\partial_\mu\phi_I-\frac{1}{2}W_\mu^A\tilde\gamma^A_{IJ}\phi_J\, ,\label{eq:Dphi4}\\
W_{A\mu\nu}&=&\partial_\mu W_{A\nu}-\partial_\nu W_{A\mu}-\tilde\epsilon_{ABC}W_{B\mu} W_{C\nu}\, .
\end{eqnarray}
The gluons are included in a slightly different manner as the color indices are not explicitly expanded by \FR or the \FA and \FC packages. This complicates the doubling of the field content necessary for the background field method. The metric for gluons is defined as:
\begin{eqnarray}
\kappa=1-4c_{HG}\frac{\phi^2}{2}-4c_{HG}^{(8)}\frac{\phi^4}{4}\, .
\end{eqnarray}
While the inverse square root expectation of $\kappa$ is explicitly entered as:
\begin{equation}\label{eq:kappainvsqrt}
\langle\kappa\rangle^{-1/2}=1+c_{HG}v_T^2+\frac{1}{2}(c_{HG}^{(8)}v_T^4+3 c_{HG}^2v_T^4)\, .
\end{equation}
This allows for the definition of the gluon field strength with explicit insertions of Eq.~\ref{eq:kappainvsqrt} such that the kinetic terms of the gluons are made canonical in the presence of $\mathcal O_{HG}$ and $\mathcal O_{HG}^{(8)}$:
\begin{equation}\label{eq:glufs}
G_{C\mu\nu}=\langle\kappa\rangle^{-1/2}\left[\partial_\mu \mathcal G_{C\nu}-\partial_\nu \mathcal G_{C\mu}-g_3\langle\kappa\rangle^{-1/2}f_{ABC}\mathcal G_{B\mu}\mathcal G_{C\nu}-g_3\langle\kappa\rangle^{-1/2}f_{ABC}\mathcal G_{B\mu}\hat {\mathcal G}_{C\nu}+\mathcal G\leftrightarrow\hat {\mathcal G}\right]\, .
\end{equation}
Where $\mathcal G$ and $\mathcal {\hat G}$ are the physical quantum and background gluon fields. Here latin indices are color indices and we have introduced the structure constants of $SU(3)_C$, $f_{ABC}$. 

Consistent with the standard usage of \FR, the unphysical fields \texttt{V[11]}, \texttt{V[12]}, and \texttt{V[13]} are introduced for the fields $W_{4\mu}$, $W_{1,2,3,\mu}$ and $G_{C\mu}=\mathcal G_{C\mu}+\hat{\mathcal G}_{C\mu}$. These are defined with Eq.~\ref{eq:weakeigenstatefields} so that the built in Feynrules covariant derivative of the fermionic fields is consistent with Eq.~4.12 of \cite{Helset:2018fgq} as well as the doubling of the field content consistent with the background field method. It should be noted that the explicit definition of the field strengths of the gauge fields implies a specific covariant derivative sign, as such the \FR parameter \texttt{FR\$DSign} should not be changed. With the above we are now able to implement the full SMEFT Lagrangian to dimension six and including class 2, 3, and 4 operators to dimension eight as well as dimension-six squared effects coming from shifts in kinetic and mass terms in the Lagrangian.

\subsection{The Bosonic Lagrangian}
From the definitions above, we are able to define the full SMEFT Lagrangian at dimension six as well as including the $\mathcal{O}\left(\frac{1}{\Lambda^4}\right)$ effects of the operators contributing to the metrics. The Class 3 operators are included in the scalar Lagrangian:
\begin{equation}\label{eq:LHiggs}
\mathcal L_{\rm Higgs}=\frac{1}{2}h_{IJ}(D_\mu\phi)_I(D_\mu\phi)_J-\frac{\lambda}{2}\left(\phi^2-v\right)^2+c_H\left(\frac{\phi^2}{2}\right)^3+c_{H}^{(8)}\left(\frac{\phi^2}{2}\right)^4\, .
\end{equation}
The substitution, 
\begin{equation}
v\to v_T-\frac{1}{4\lambda}\left(\frac{3}{2}c_H v_T^3-c_{H}^{(8)}v_T^5\right)-\frac{9}{128\lambda^2}c_H^2v_T^5+\mathcal{O}\left(\frac{1}{\Lambda^6}\right)\, ,
\end{equation}
is then made to write the Higgs potential in terms of the true vacuum expectation value, defined as the vacuum expectation value which minimizes the potential. The class four operators are then included through the gauge boson Lagrangians:
\begin{equation}
\mathcal L_{\rm Gauge}=-\frac{1}{4}g_{AB}W_{A\mu\nu}W_{B\mu\nu}-\frac{1}{4}\kappa G_{C\mu\nu}G_{C\mu\nu}\, .
\end{equation}
In order to gauge fix the quantum fields we define the following quantities:
\begin{eqnarray}
\mathcal G_X^{\rm Weak}&=&\partial_\mu W_{X\mu}-\tilde\epsilon_{XCD}\hat W_{C\mu}W_{D\mu}+\frac{\xi}{2}\hat g_{XC}\phi_I\hat h_{IK}\tilde\gamma_{CKJ}\hat \phi_J\, ,\\
\mathcal G_A^{\rm Color}&=&\langle\kappa\rangle^{-1/2}\partial_\mu G_{\mu A}-g_3\langle\kappa\rangle^{-1}f_{ABC}\hat G_{\mu B}G_{\mu C}\, .
\end{eqnarray}
Here hatted quantities correspond to fields and metrics containing only the background fields and unhatted quantities should be understood as the quantum fields. This is true only in the gauge fixing and ghost Lagrangians. Elsewhere the background and quantum fields are treated the same. Again the implementation of the gluons is handled differently from the weak gauge bosons as explained in Subsection~\ref{sub:fieldsmetrics}. The gauge fixing Lagrangian is then defined as:
\begin{equation}\label{eq:GF}
\mathcal L_{\rm GF}=-\frac{\hat g_{AB}}{2\xi}\mathcal G_A^{\rm Weak}\mathcal G_B^{\rm Weak}-\frac{\hat \kappa}{2\xi_G}\mathcal G_A^{\rm Color}\mathcal G_A^{\rm Color}\, .
\end{equation}
The ghost Lagrangians contain more terms, and as such they were written term by term. For the ghosts associated with the weak gauge bosons we have:
\begin{eqnarray}
\mathcal L_{\rm a}&=&-\hat g_{AB}\bar u_B\partial^2 u_A\, ,\\
\mathcal L_{\rm b}&=& -\partial_\mu (\hat g_{AB}\bar u_B) u_C\tilde\epsilon_{ADC}(W_{D\mu}+\hat W_{D\mu})\, ,\\
\mathcal L_{\rm c}&=&\hat g_{AB}\bar u_B(\partial_\mu\bar u_C)\tilde\epsilon_{ADC}\hat W_{D\mu}\, ,\\
\mathcal L_{\rm d}&=&-\hat g_{AB}\bar u_B u_C\tilde\epsilon_{ADE}\tilde\epsilon_{EFC}\hat W_{D\mu}(W_{F\mu}+\hat W_{F\mu})\, ,\\
\mathcal L_{\rm scalar}&=&-\frac{\xi}{4}\hat g_{AB}\hat g_{AD}^{-1}\bar u_Bu_C\phi_J\tilde\gamma_{CIJ}\hat h_{IK}\tilde\gamma_{DKL}\hat\phi_L\, ,\\
\mathcal L_{\rm ghost}^{\rm Weak}&=&\mathcal L_{\rm a}+\mathcal L_{\rm b}+\mathcal L_{\rm c}+\mathcal L_{\rm d}+\mathcal L_{\rm scalar}\, .
\end{eqnarray}
While for the gluons we have:
\begin{eqnarray}
\mathcal L_{\rm a}^G&=&\hat\kappa\langle\kappa\rangle^{-1}u^G_A\partial^2 u^G_A\, ,\\
\mathcal L_{\rm b}^G&=&\langle\kappa\rangle^{-3/2}g_3f_{ADC}\partial_\mu (\hat\kappa u^G_A)u^G_C(G_{\mu D}+\hat G_{\mu D})\, ,\\
\mathcal L_{\rm c}^G&=&-\hat\kappa\langle\kappa\rangle^{-3/2}g_3f_{ADC}u^G_A(\partial_\mu u_C^G)\hat G_{\mu D}\, ,\\
\mathcal L_{\rm d}^G&=&\hat\kappa\langle\kappa\rangle^{-2}g_3^2f_{ADE}f_{EFC}u^G_Au^G_C\hat G_{\mu D}(G_{\mu F}+\hat G_{\mu F})\, ,\\
\mathcal L_{\rm ghost}^G&=&\mathcal L_{\rm a}^G+\mathcal L_{\rm b}^G+\mathcal L_{\rm c}^G+\mathcal L_{\rm d}^G\, .
\end{eqnarray}
In \cite{Dekens:2019ept} the authors found a sign convention discrepancy between the standard ghost term and that of \cite{Helset:2018fgq}. This sign is included explicitly in the above Lagrangians.

The only remaining purely bosonic operators are of class 1. They are defined as:
\begin{equation}
\mathcal{L}_{\rm cl1}=c_G f_{ABC}G_{A\mu\nu}G_{B\nu\rho}G_{C\rho\mu}+c_W\epsilon_{ABC}W_{A\mu}W_{B\nu\rho}W_{C\rho\mu}\, .
\end{equation}

\subsection{Implementation of the fermionic operators}
As mentioned below Eq.~\ref{eq:glufs}, this \FR package make use of unphysical fields associated with the SM gauge group in order to use the internal definition of the covariant derivative. As such the fermion gauge-kinetic terms are trivially defined as,
\begin{equation}
\mathcal L_{\rm Fermions}=i\bar Q\slashed D Q+i\bar L \slashed D L+i\bar u_R\slashed D u_R+i\bar d_R\slashed D d_R+i\bar l_R\slashed D l_R\, ,
\end{equation}
where there is an implicit sum over flavors in the above expression. The CKM matrix is implemented in the same manner as with the SM \FR file \cite{Alloul:2013bka}, and is defined below in Eq.~\ref{eq:CKM}. 

The Yukawa couplings plus the class 5 operator contributions can be summed as:
\begin{eqnarray}
\mathcal L_{\rm Yukawa}+\mathcal L_{\rm cl5}&=&-\left[y_d-c_{dH}(H^\dagger H)\right]_{ij}\bar Q_i d_{Rj} H-\left[y_u-c_{uH}(H^\dagger H)\right]_{ij} \bar Q_iu_{Rj}\tilde H\nonumber\\
&&-\left[y_l-c_{eH}(H^\dagger H)\right]_{ij}\bar L l_R H+h.c.\, ,
\end{eqnarray}
We rotate between the weak (unprimed) and the mass eigenstate (primed) bases via the definitions:
\begin{equation}
\begin{array}{cccc}
u_L=\mathcal U_{uL}u'_L\, ,\ \ \ \ \ \ \ \ \ \ &u_R=\mathcal U_{uR}u'_R\, ,\ \ \ \ \ \ \ \ \ \ &\nu_L=\mathcal U_{\nu L}\nu'_L\, ,\ \ \ \ \ \ \ \ \ \ &\\
d_L=\mathcal U_{dL}d'_L\, ,\ \ \ \ \ \ \ \ \ \ &d_R=\mathcal U_{dR}d'_R\, ,\ \ \ \ \ \ \ \ \ \ &e_L=\mathcal U_{eL}e'_L\, ,\ \ \ \ \ \ \ \ \ \ &e_R=\mathcal U_{eR}e'_R\, .\ \ \ \ \ \\
\end{array}
\end{equation}
where flavor indices are suppressed. The definitions of the rotation of the left-handed up and down fields imply the definition of the CKM matrix,
\begin{equation}\label{eq:CKM}
V^{CKM}=\mathcal U^\dagger_{uL}\mathcal U_{dL}\, .
\end{equation}
This allows us to write the above Lagrangian instead in terms of barred masses, the CKM matrix, and rotated Wilson coefficients. For example, for the down quarks:
\begin{eqnarray}
\left.\mathcal L_{\rm Yukawa}+\mathcal L_{\rm cl5}\right|_d&=&-\left[y_d-c_{dH}\frac{v^2}{2}\right]_{ij}\bar Q_i d_{Rj} H+c_{dH,ij}(H^\dagger H-v^2/2)\bar Q_i d_{Rj} H\\
&\equiv&-V^{CKM}_{ij}\left(\frac{\sqrt{2}\bar M_d}{v}\right)_{jj}\bar u'_{L,i}d'_{R,j}H_1+\left(\frac{\sqrt{2}\bar M_d}{v}\right)_{ii}\bar d'_{L,i}d'_{R,i}H_2\nonumber\\
&&+\left(H^\dagger H-\frac{v^2}{2}\right)\left[V^{CKM}_{ik}c'_{dH,kl}\bar u'_{Li} d'_{Rj} H_1+c'_{dH,ij}\bar d'_{Li} d'_{Rj} H_2\right]\, ,
\end{eqnarray}
where $c'_{dH}\equiv\mathcal U^\dagger_{dL}c_{dH}\mathcal U_{dR}$ and $H_{1,2}$ correspond to the first and second components of the doublet defined in Eq.~\ref{eq:doublet}.  

By choice, no other fermionic operators are rotated in this manner. The CKM matrix is also defined only as above, and does not take into account any dimension-six effects. For further details on flavor and the CKM matrix in the SMEFT see, for example, \cite{Descotes-Genon:2018foz,Faroughy:2020ina,Brivio:2017btx}.

The remaining fermionic operators and corresponding Lagrangians at dimension-six are adopted directly from the \SSim package \cite{Brivio:2017btx}. In many cases the fields are redefined so that indices are explicitly expanded, greatly speeding up the calculation of the Lagrangian in Feynrules. This has the consequence that class relationships cannot be used in Feynarts and that each fermion and its interactions are defined separately in the Feynarts model file instead of grouping flavors together. In order to define the Class 7 operators we have made use of the following identities \cite{Helset:2020yio}:
\begin{eqnarray}
(H^\dagger i \overleftrightarrow{D}_\mu H)&=&-\frac{1}{g_1}\phi_I\tilde\gamma_{4IJ}(D_\mu\phi)_J\, ,\\
(H^\dagger i \overleftrightarrow{D}^A_\mu H)&=&-\frac{1}{g_2}\phi_I\tilde\gamma_{AIJ}(D_\mu\phi)_J\, ,\ \ \ \ \ {\rm for\ }A=1,2,3\, .
\end{eqnarray}
Given the Lagrangians defined above, the full Lagrangian which may be fed into Feynrules is given by:
\begin{equation}\label{eq:fullL}
\mathcal L_{\rm Full}=\mathcal L_{\rm Higgs}+\mathcal L_{\rm Gauge}+\mathcal L_{\rm GF}+\mathcal L_{\rm ghost}^{\rm Weak}+\mathcal L_{\rm ghost}^G+\mathcal L_{\rm cl1}+\mathcal L_{\rm Fermions}+\mathcal L_{\rm Yukawa}+\mathcal L_{\rm cl5}+\mathcal L_{\rm cl6}+\mathcal L_{\rm cl7}+\mathcal L_{\rm cl8}\, .
\end{equation}

Consistency at one-loop in the SMEFT requires the subtraction of the terms linear in the quantum fields from this Lagrangian through appropriate choice of the sources of the quantum fields \cite{thooft,Abbott:1981ke}. This is not done for the Lagrangian above, but is effectively implemented when considering only one-particle irreducible diagrams and by restricting background fields to be the only external field lines.

\section{Ward Identities}\label{sec:ward}
In a companion paper \cite{Corbett:2020ymv}, the Ward identities for the SMEFT, first presented in \cite{Corbett:2019cwl}, are verified for single insertions of the dimension-six operator coefficients. In this section we further confirm the power of the Feynrules package presented here by deriving the dimension-six squared contributions coming from Class 2, 3, and 4 operators of Table~\ref{tab:op59} in Appendix~\ref{app:ops}. This verification serves two purposes, to demonstrate the consistent inclusion of shifts to kinetic and mass terms coming from these operators, and to demonstrate the breakdown of the Ward Identities, and therefore the gauge invariance, for calculations when these contributions are not included in amplitudes which contain dimension-six squared terms inconsistently. 

When calculating an amplitude to $\mathcal O(1/\Lambda^4)$, the amplitude and amplitude squared are formally defined as:
\begin{eqnarray}
\mathcal A&=&\mathcal A_{\rm SM}+\frac{c^{(6)}}{\Lambda^2}\mathcal A_6+\frac{(c^{(6)})^2+c^{(8)}}{\Lambda^4}\mathcal A_8+\mathcal O\left(\frac{1}{\Lambda^6}\right)\, ,\\
|\mathcal A|^2&=&|\mathcal A_{\rm SM}|^2+\frac{c^{(6)}}{\Lambda^2}2{\rm Re}[\mathcal A_6^* \mathcal A_{\rm SM}]+\frac{(c^{(6)})^2+c^{(8)}}{\Lambda^4}2{\rm Re}[\mathcal A_8^*\mathcal A_{\rm SM}]+\frac{(c^{(6)})^2}{\Lambda^4}|\mathcal A_6|^2+\mathcal O\left(\frac{1}{\Lambda^6}\right)\, .\nonumber\\ 
\end{eqnarray}
It is common, however, to drop the $c^{(8)}$ term on the grounds that most studies exclude dimension-eight operators. However, the $(c^{(6)})^2$ terms are frequently kept either for convenience or as an estimate of the theory error or convergence of the SMEFT series\footnote{This is technically problematic already, as inclusion of $(c^{(6)})^2$ terms without $c^{(8)}$ clearly neglects a contribution of the same order, $1/\Lambda^4$. A more subtle problem is that the error associated with the definition of a non-redundant basis at dimension-six is $\mathcal O(1/\Lambda^4)$ \cite{Politzer:1980me,Georgi:1991ch,Arzt:1993gz,Simma:1993ky}. This issue is discussed further in \cite{Hays:2020scx}.}. Frequently, however, the $(c^{(6)})^2\mathcal A_8$ term is neglected, but the square of $c^{(6)}\mathcal A_6$ is included. Worse, in many instances the $(c^{(6)})^2\mathcal A_8$ is partially included, where direct insertions of operator effects in the amplitude are included, but the dimension-six squared shifts to field normalizations and masses are neglected. This is clearly inconsistent from a power counting perspective, however, in verifying the Ward identities below, we demonstrate this also violates gauge invariance. 

\subsection{Conventions}
We follow the same conventions of \cite{Corbett:2020ymv}. The two point functions are defined as the double variation of the effective action (in this case the full Lagrangian of Eq.~\ref{eq:fullL}):
\begin{eqnarray}
-i \Gamma^{\hat{V},\hat{V}'}_{\mu \nu}(k)&=&
\left(-g_{\mu \nu} k^2 + k_\mu k_\nu + g_{\mu \nu} \bar{M}_{\hat{V}}^2\right)\delta^{\hat{V} \hat{V}'}
+\left(-g_{\mu \nu} +\frac{k_\mu k_\nu}{k^2}  \right) \Sigma_{T}^{\hat{V},\hat{V}'}- \frac{k_\mu k_\nu}{k^2}
\Sigma_{L}^{\hat{V},\hat{V}'},\\
\frac{\delta^2 \Gamma}{{\delta  \hat{\Phi}^{3} \delta \hat{\mathcal{A}}^{3 \nu}}} &=&
{-\frac{\delta^2 \Gamma}{\delta \hat{\mathcal{A}}^{3 \nu}\delta  \hat{\Phi}^{3}}} =
 i k^\nu \left[i \,\bar{M}_{\mathcal{Z}}+ \Sigma^{\hat{Z}\hat{\chi}}(k^2) \right], \\
\frac{\delta^2 \Gamma}{\delta  \hat{\Phi}^{3} \delta  \hat{\Phi}^{3}}  &=& i k^2 + i \Sigma^{\hat{\chi}\hat{\chi}}(k^2),\\
\frac{\delta^2 \Gamma}{{\delta  \hat{\Phi}^{\pm} \delta \hat{\mathcal{W}}^{\mp \nu}}} &=&
= {-i} \, k^\nu \left[{\pm} \,\bar{M}_W + \Sigma^{{\hat{\Phi}^\pm \hat{\mathcal{W}}^\mp}}(k^2) \right], \\
{\frac{\delta^2 \Gamma}{\delta \hat{\mathcal{W}}^{\pm \nu} \delta  \hat{\Phi}^{\mp}}}&=&
{i  \, k^\nu \left[\mp \,\bar{M}_W + \Sigma^{{\hat{\mathcal{W}}^\pm \hat{\Phi}^\mp}}(k^2) \right]},\\
\frac{\delta^2 \Gamma}{\delta  \hat{\Phi}^{+} \delta  \hat{\Phi}^{-}}  &=& i k^2 + i \Sigma^{\hat{\Phi}^+\hat{\Phi}^-}(k^2),\\
\frac{\delta\Gamma}{\delta\hat H}&=&iT^H.
\end{eqnarray}
With these definitions it was found that the Ward identities take the simple form:
\begin{eqnarray}\label{eq:wardids}
0&=&\Sigma_L^{\hat {\mathcal A}\hat {\mathcal A}},\label{eq:WIAA}\\
0&=&\Sigma_L^{\hat {\mathcal A}\hat {\mathcal Z}},\label{eq:WIAZ}\\
0 &=& \Sigma_L^{\hat{{\mathcal W}^\pm} \hat{{\mathcal W}}^\mp}(k^2) \pm \bar{M}_W \Sigma^{{\hat{\Phi}^\pm \hat{{\mathcal W}}^\mp}}(k^2),\label{eq:WIWW} \\
0 &=& k^2 \Sigma^{\hat{{\mathcal W}}^\pm \hat{\Phi}^\mp}(k^2)\pm \bar{M}_W \, \Sigma^{{\hat{\Phi}^\pm  \hat{\Phi}^\mp}}(k^2)
\mp \frac{\bar{g}_2}{4} T^H \sqrt{h}_{44}\left(\sqrt{h}^{11}+\sqrt{h}^{22}\right)\label{eq:WIWphi}\\
0 &=& \Sigma_L^{\hat{{\mathcal Z}}\hat{{\mathcal Z}}}(k^2)- i \bar{M}_{{ Z}} \Sigma^{\hat{{\mathcal Z}} \hat{\chi}}(k^2),\label{eq:WIZZ} \\
0 &=& k^2 \Sigma^{\hat{{\mathcal Z}} \hat{\chi}}(k^2)- i \bar{M}_{{Z}} \, \Sigma^{\hat{\chi} \hat{\chi}}(k^2) + i \, \frac{\bar{g}_Z}{2}\sqrt{h}_{44}\sqrt{h}^{33} T^H ,\label{eq:WIZchi}
\end{eqnarray}
Where $\sqrt{h}_{ii}$ is the square-root of the metric expectation $\langle h\rangle$ and $\sqrt{h}^{ii}$ is the inverse-square-root of the metric expectation, as defined in Eqs.~\ref{eq:expmetrics}~and~\ref{eq:metricinvsqrts}. We have dropped the off-diagonal contributions from the metric $h$, as they are vanishing up to and including $\mathcal O(1/\Lambda^4)$.

\subsection{Verification of Ward Identities for dimension-six squared contributions}\label{sec:verifyWI}
The full set of amplitudes and their contributions from each combination of Wilson coefficients are tabulated in Appendix~\ref{app:amps}. For simplicity we only consider the UV divergences of these amplitudes. Here we look at some special cases, confirming the Ward identities and demonstrating that they breakdown when the full contribution of the shifts of the fields are not included. Table~\ref{tab:WIviolationsummary} shows the combinations of Wilson coefficients which violate each of the Ward identities in Eqs~\ref{eq:WIAA}--\ref{eq:WIZchi} when field shifts are not consistently kept up to $\mathcal O(1/\Lambda^4)$. 

Appendix~\ref{app:amps} shows explicit dependence on the shifts of the fields in the amplitudes for $c_i^2$ in the form of a parameter $\Delta$. The amplitude in the absence of the $O(1/\Lambda^4)$ field shifts can be obtained by $\Delta\to0$ and the correct amplitude by $\Delta\to1$. The $\Delta$ dependence is only included for the cases of $c_i^2$ and not for $c_ic_j$ with $i\ne j$ for brevity. In the discussion below we will refer to the $\Delta$ dependence for amplitudes depending on $c_i^2$, but also include information on the breakdown of the Ward identities when the the field shifts are not included consistently for $c_ic_j$ dependence even though the $\Delta$ dependence is not included in the Appendix.

\begin{table}
\begin{tabular}{|l|l|}
\hline
\textbf{Ward Identity}&\textbf{Wilson coefficient combination}\\
\hline
\hline
Equation~\ref{eq:WIAA} \hfill(QED)&$c_{HW}^2$\\
\hline
Equation~\ref{eq:WIAZ} \hfill($ZA$)&$c_{HB}^2,\ c_{HW}^2,\ c_{HB}c_{HWB},\ c_{HW}c_{HWB}$\\
\hline
Equation~\ref{eq:WIWW} \hfill$(\Sigma^{\mathcal W\mathcal W}_L+\cdots)$&$c_{H\Box}^2,\ c_{HD}^2,\ c_{HW}^2,\ c_{HWB}^2,\ c_{H\Box}c_{HD}$\\
\hline
Equation~\ref{eq:WIWphi} \hfill$\ \ \ \ \ (k^2\Sigma^{\mathcal W\phi}+\cdots)$&$c_{H\Box}^2,\ c_{HD}^2,\ c_{HB}^2,\ c_{HW}^2,\ c_{HWB}^2,\ c_{H\Box}c_{HD},\ c_{HB}c_{HWB}$\\
\hline
Equation~\ref{eq:WIZZ} \hfill$(\Sigma^{\mathcal Z\mathcal Z}_L+\cdots)$&$c_{H\Box}^2,\ c_{HD}^2,\ c_{HB}^2,\ c_{HW}^2,\ c_{HWB}^2,\ c_{H\Box}c_{HD},$\\
&$c_{HB}c_{HWB},\ c_{HW}c_{HWB}$\\
\hline
Equation~\ref{eq:WIZchi} \hfill$(k^2\Sigma^{\mathcal Z\chi}+\cdots)$&$c_{H\Box}^2,\ c_{HD}^2,\ c_{HB}^2,\ c_{HW}^2,\ c_{HWB}^2,\ c_{H\Box}c_{HD},$\\
&$c_{HB}c_{HWB},\ c_{HW}c_{HWB}$\\
\hline
\end{tabular}
\caption{A summary table of the Ward Identities which are violated for each combination of Wilson coefficients, $c_ic_j$, when field redefinitions are not properly maintained to $\mathcal O(1/\Lambda^4)$. The combinations, $c_Hc_i$ and $c_{HD}c_i$ for all $c_i$, $c_{H\Box}c_{HB}$, $c_{H\Box}c_{HW}$, $c_{H\Box}c_{HWB}$, $c_{HB}c_{HW}$, do not violate the Ward identities when field redefinitions handled inconsistently. In total one-third of the possible combinations break at least one of the Ward identities.}\label{tab:WIviolationsummary}
\end{table}

\subsubsection{Longitudinal photons and $\mathcal Z\gamma$ mixing}
Inconsistently expanding to $\mathcal O(1/\Lambda^2)$ in the worst case generates a longitudinal component of the photon field. In Eq.~\ref{eq:cHW2} it is demonstrated that when the field shifts are not properly included (i.e. $\Delta\to0$ in this equation) that a longitudinal component of the photon, proportional to $c_{HW}^2$ is generated. This clearly is a violation of the gauge symmetry of the SM and indeed breaks $U(1)_{QED}$ which is meant to be preserved after EWSB.
In addition longitudinal mixing between the $\mathcal Z$ and photon can be generated. This can be seen explicitly in Eqs~\ref{eq:cHB2}~and~\ref{eq:cHW2}, i.e. with contributions proportional to $c_{HB}^2$ and $c_{HW}^2$. This is also the case for $c_{HB}c_{HWB}$ and $c_{HW}c_{HWB}$, which is not shown explicitly in the appendix.

\subsubsection{Charged Ward identities}
For the more complex ward identities we need to include dependence of $\bar M_{W,Z}$ and $\bar g_{2,Z}$ in the verification of the Ward identities. The definition of these barred quantities can be found in Appendix~\ref{app:bars}. In the trivial case that $\bar M_W$ and $\bar g_2$ don't depend on a given Wilson coefficient the Ward identities remain simple, for example, considering $c_{HB}^2$ dependence we have:
\begin{eqnarray}
0&=&\Sigma_L^{\hat{\mathcal W}^\pm\hat{\mathcal W}^\mp}\pm\bar M_W\Sigma^{\hat\Phi^\pm\hat{\mathcal W}^\mp}\nonumber\\[8pt]
&\to&\Sigma_{L,c_{HB}^2}^{\hat{\mathcal W}^\pm\hat{\mathcal W}^\mp}\pm \frac{g_2v_T}{2}\Sigma^{\hat\Phi^\pm\hat{\mathcal W}^\mp}_{c_{HB}^2}\, ,
\end{eqnarray}
\begin{eqnarray}
0&=&k^2 \Sigma^{\hat{{\mathcal W}}^\pm \hat{\Phi}^\mp}\pm \bar{M}_W \, \Sigma^{{\hat{\Phi}^\pm  \hat{\Phi}^\mp}}
\mp \frac{\bar{g}_2}{4} T^H \sqrt{h}_{44}\left(\sqrt{h}^{11}+\sqrt{h}^{22}\right)\nonumber\\[8pt]
&\to&k^2 \Sigma^{\hat{{\mathcal W}}^\pm \hat{\Phi}^\mp}_{c_{HB}^2}\pm \frac{g_2v_T}{2} \, \Sigma^{{\hat{\Phi}^\pm  \hat{\Phi}^\mp}}_{c_{HB}^2}
\mp \frac{\bar{g}_2}{4} T^H_{c_{HB}^2}\, .
\end{eqnarray}
Where $\Sigma_{c_ic_j}$ and $T_{c_ic_j}$ represent the dependence of $\Sigma$ and $T$ on the combination $c_ic_j$. These identities are satisfied in a straightforward manner by the form of the amplitudes in Eq.~\ref{eq:cHB2} for $\Delta\to1$. If, however, we take $\Delta\to0$ we find the second Ward identity is not preserved:
\begin{equation}
k^2 \Sigma^{\hat{{\mathcal W}}^\pm \hat{\Phi}^\mp}\pm \frac{g_2v_T}{2} \, \Sigma^{{\hat{\Phi}^\pm  \hat{\Phi}^\mp}}
\mp \frac{\bar{g}_2}{4} T^H=-c_{HB}^2\frac{g_1^2g_2v_T^7}{512\pi^2\epsilon}[3g_1^2\xi^2-8g_2^2(-3+\xi)]\ne0\, .
\end{equation}

 If we look instead at the more complex case of $c_{HB}c_{HW}$ dependence we find, because $\bar M_W$ and $\bar g_2$ depend on $c_{HW}$:
\begin{eqnarray}
0&=&\Sigma_L^{\hat{\mathcal W}^\pm\hat{\mathcal W}^\mp}\pm\bar M_W\Sigma^{\hat\Phi^\pm\hat{\mathcal W}^\mp}\nonumber\\
&\to&\Sigma_L^{\hat{\mathcal W}^\pm\hat{\mathcal W}^\mp}\pm \left[\frac{g_2v_T}{2}\Sigma_{c_{HW}c_{HB}}^{\hat\Phi^\pm\hat{\mathcal W}^\mp}(k^2)+\frac{g_2v_T^3}{2}c_{HW}\Sigma_{c_{HB}}^{\hat\Phi^\pm\hat{\mathcal W}^\mp}\right]\, ,\label{eq:chargedcHBcHW1}
\end{eqnarray}
\begin{eqnarray}
0&=&k^2 \Sigma^{\hat{{\mathcal W}}^\pm \hat{\Phi}^\mp}\pm \bar{M}_W \, \Sigma^{{\hat{\Phi}^\pm  \hat{\Phi}^\mp}}
\mp \frac{\bar{g}_2}{4} T^H \sqrt{h}_{44}\left(\sqrt{h}^{11}+\sqrt{h}^{22}\right)\nonumber\\
&\to&k^2 \Sigma^{\hat{{\mathcal W}}^\pm \hat{\Phi}^\mp}\pm \left[\frac{g_2v_T}{2} \, \Sigma_{c_{HB}c_{HW}}^{{\hat{\Phi}^\pm  \hat{\Phi}^\mp}}+\frac{g_2v_T^3}{2}c_{HW} \, \Sigma_{c_{HB}}^{{\hat{\Phi}^\pm  \hat{\Phi}^\mp}}\right]
\mp \left[\frac{{g}_2}{4} T^H_{c_{HB}c_{HW}}+\frac{{g}_2v_T^2}{4}c_{HW} T^H_{c_{HB}}\right]\, .\nonumber\\
\label{eq:chargedcHBcHW2}
\end{eqnarray}
In order to confirm these Ward identities we need the appropriate $c_{HB}$ dependence of each amplitude, they can be found in \cite{Corbett:2020ymv}:
\begin{eqnarray}
\Sigma_{c_{HB}}^{\hat\Phi^\pm\hat{\mathcal W}^\mp}&=&c_{HB}\frac{g_1^2g_2v_T^3}{64\pi^2\epsilon}(3+\xi)\, ,\\
\Sigma_{c_{HB}}^{{\hat{\Phi}^\pm  \hat{\Phi}^\mp}}&=&c_{HB}\frac{g_1^2v_T^2}{128\pi^2\epsilon}[v_T^2(9g_1^2+9g_2^2+4\lambda\xi)-4(3+\xi)k^2]\, ,\\
T^H_{c_{HB}}&=&c_{HB}\frac{g_1^2v_T^5}{128\pi^2\epsilon}[9g_1^2+9g_2^2+4\lambda\xi]\, .
\end{eqnarray}
With these amplitudes and those of Eq.~\ref{eq:cHBcHW} we see Eqs.~\ref{eq:chargedcHBcHW1}~and~\ref{eq:chargedcHBcHW2} are indeed satisfied. In this case the Ward identities are satisfied even when the shifts in the fields are neglected at $\mathcal O(1/\Lambda^2)$.

Another interesting feature of the Ward identities to inspect is the explicit dependence of Eq.~\ref{eq:WIWphi} on $\sqrt{h}$. Considering the $c_{H\Box}c_{HD}$ dependence the form of this Ward identity becomes:
\begin{eqnarray}
0&=&k^2 \Sigma^{\hat{{\mathcal W}}^\pm \hat{\Phi}^\mp}\pm \bar{M}_W \, \Sigma^{{\hat{\Phi}^\pm  \hat{\Phi}^\mp}}
\mp \frac{\bar{g}_2}{4} T^H \sqrt{h}_{44}\left(\sqrt{h}^{11}+\sqrt{h}^{22}\right)\nonumber\\[8pt]
&\to&k^2 \Sigma_{c_{H\Box}c_{HD}}^{\hat{{\mathcal W}}^\pm \hat{\Phi}^\mp}\pm \frac{g_2v_T}{2} \, \Sigma_{c_{H\Box}c_{HD}}^{{\hat{\Phi}^\pm  \hat{\Phi}^\mp}}
\mp \frac{{g}_2}{2} \left[T^H_{c_{H\Box}c_{HD}}-c_{H\Box}v_T^2T^H_{c_{HD}}+\frac{c_{HD}v_T^2}{4}T^H_{c_{H\Box}}+\frac{c_{H\Box}c_{HD}v_T^4}{4}T^H_{\rm SM}\right]\, .\nonumber\\[8pt]
\end{eqnarray}
Where $T^H_{\rm SM}$ is the SM contribution to the tadpole. In order to verify this identity we need the following amplitudes:
\begin{eqnarray}
T^H_{\rm SM}&=&\frac{v_T^3}{256\pi^2\epsilon}[3g_1^4+9g_2^4+96\lambda^2+12g_2^2\lambda\xi+g_1^2(6g_2^2+4\lambda\xi)]\, ,\\
T^H_{c_{H\Box}}&=&c_{H\Box}\frac{v_T^5}{256\pi^2\epsilon}[3g_1^4+9g_2^4+608\lambda^2+12g_2^2\lambda\xi+g_1^2(6g_2^2+4\lambda\xi)]\, ,\\
T^H_{c_{HD}}&=&c_{HD}\frac{v_T^5}{1024\pi^2\epsilon}[15g_1^4+9g_2^4-608\lambda^2-12g_2^2\lambda\xi+g_1^2(30g_2^2-4\lambda\xi)]\, ,
\end{eqnarray}
We see the Ward identity is indeed satisfied given Eq.~\ref{eq:cHBoxcHD}. However, if $\Delta$ is taken to be zero, we instead find:
\begin{eqnarray}
k^2 \Sigma_{c_{H\Box}c_{HD}}^{\hat{{\mathcal W}}^\pm \hat{\Phi}^\mp}&\pm& \frac{g_2v_T}{2} \, \Sigma_{c_{H\Box}c_{HD}}^{{\hat{\Phi}^\pm  \hat{\Phi}^\mp}}
\mp \frac{{g}_2}{2} \left[T^H_{c_{H\Box}c_{HD}}-c_{H\Box}v_T^2T^H_{c_{HD}}+\frac{c_{HD}v_T^2}{4}T^H_{c_{H\Box}}+\frac{c_{H\Box}c_{HD}v_T^4}{4}T^H_{\rm SM}\right]\nonumber\\
&=&c_{H\Box}c_{HD}\frac{g_2v_T^7}{1024\pi^2\epsilon}[3g_1^4-96\lambda^2+44g_2^2\lambda\xi+g_1^2(6g_2^2+4\lambda\xi)-g_2^4(-9+\xi^2)]\ne0\, .\nonumber\\
\end{eqnarray}
We refrain from expanding every combination $c_ic_j$ here, and stop with these examples. The validity of the Ward identities for the remaining Wilson coefficient combinations has been confirmed for all charged Ward identities using the amplitudes enumerated in Appendix~\ref{app:amps} and \cite{Corbett:2020ymv}.

\subsubsection{Neutral Ward identities involving $\hat{\mathcal Z}$ and $\hat\chi$}
Just as in the case of the charged ward identities we may proceed with Eqs~\ref{eq:WIZZ}~and~\ref{eq:WIZchi}. We restrict ourselves to the example of $c_{HD}^2$ as it contributes to $\bar M_Z$ and $\sqrt{h}$:
\begin{eqnarray}
0 &=& \Sigma_L^{\hat{{\mathcal Z}}\hat{{\mathcal Z}}}- i \bar{M}_{{ Z}} \Sigma^{\hat{{\mathcal Z}} \hat{\chi}},\label{eq:WIZZ}\nonumber\\
&\to&\Sigma_{L,c_{HD}^2}^{\hat{{\mathcal Z}}\hat{{\mathcal Z}}}(k^2)-i\frac{\sqrt{g_1^2+g_2^2}v_T}{2}\left[\Sigma^{\hat{{\mathcal Z}} \hat{\chi}}_{c_{HD}^2}+\frac{c_{HD}v_T^2}{4}\Sigma^{\hat{{\mathcal Z}} \hat{\chi}}_{c_{HD}}-\frac{c_{HD}^2v_T^4}{32}\Sigma^{\hat{{\mathcal Z}} \hat{\chi}}_{\rm SM}\right]\, ,\\
0 &=& k^2 \Sigma^{\hat{{\mathcal Z}} \hat{\chi}}- i \bar{M}_{{Z}} \, \Sigma^{\hat{\chi} \hat{\chi}} + i \, \frac{\bar{g}_Z}{2}\sqrt{h}_{44}\sqrt{h}^{33} T^H\,  \nonumber\\
&=&k^2 \Sigma_{c_{HD}^2}^{\hat{{\mathcal Z}} \hat{\chi}}-i\frac{\sqrt{g_1^2+g_2^2}v_T}{2}\left[\Sigma^{\hat{\chi} \hat{\chi}}_{c_{HD}^2}+\frac{c_{HD}v_T^2}{4}\Sigma^{\hat{\chi} \hat{\chi}}_{c_{HD}}-\frac{c_{HD}^2v_T^4}{32}\Sigma^{\hat{\chi} \hat{\chi}}_{\rm SM}\right]+ i\frac{\sqrt{g_1^2+g_2^2}}{2}T^H_{c_{HD}^2}\, .\nonumber\\
\end{eqnarray}
To confirm the validity of these ward identities we need the following amplitudes:
\begin{eqnarray}
\Sigma^{\hat{{\mathcal Z}} \hat{\chi}}_{\rm SM}&=&-\frac{iv_T}{128\pi^2\epsilon}\sqrt{g_1^2+g_2^2}(g_1^2+3g_2^2)(3+\xi)\, ,\\
\Sigma^{\hat{{\mathcal Z}} \hat{\chi}}_{c_{HD}}&=&-c_{HD}\frac{3iv_T^3}{512\pi^2\epsilon}\sqrt{g_1^2+g_2^2}[16\lambda+g_1^2(5+\xi)+g_2^2(11+3\xi)]\, ,\\
\Sigma^{\hat{\chi} \hat{\chi}}_{\rm SM}&=&\frac{1}{256\pi^2\epsilon}[v_T^2(3g_1^4+9g_2^4+96\lambda^2+12g_2^2\lambda\xi+g_1^2[6g_2^2+4\lambda\xi])-4(g_1^2+3g_2^2)(3+\xi)k^2]\, ,\\
\Sigma^{\hat{\chi} \hat{\chi}}_{\rm c_{HD}}&=&c_{HD}\frac{v_T^2}{256\pi^2\epsilon}[v_T^2(3g_1^2-2\lambda[88\lambda+3g_2^2\xi]+g_1^2[6g_2^2-2\lambda\xi])-2(24\lambda+3g_2^2[4+\xi]+g_1^2[6+\xi])k^2]\, ,\nonumber\\
\\
T^H_{c_{HD}}&=&c_{HD}\frac{v_T^5}{1024\pi^2\epsilon}[15g_1^4+9g_2^4-608\lambda^2-12g_2^2\lambda\xi+g_1^2(30g_2^2-4\lambda\xi)]\, .
\end{eqnarray}
Combined with Appendix~\ref{app:amps} we find these Ward identities are indeed satisfied. If, however, the $\mathcal O(1/\Lambda^4)$ field shifts are not included we find:
\begin{eqnarray}
\Sigma_{L,c_{HD}^2}^{\hat{{\mathcal Z}}\hat{{\mathcal Z}}}(k^2)&-&i\frac{\sqrt{g_1^2+g_2^2}v_T}{2}\left[\Sigma^{\hat{{\mathcal Z}} \hat{\chi}}_{c_{HD}^2}+\frac{c_{HD}v_T^2}{4}\Sigma^{\hat{{\mathcal Z}} \hat{\chi}}_{c_{HD}}-\frac{c_{HD}^2v_T^4}{32}\Sigma^{\hat{{\mathcal Z}} \hat{\chi}}_{\rm SM}\right]\nonumber\\
&=&c_{H\Box}c_{HD}\frac{g_2v_T^7}{1024\pi^2\epsilon}[3g_1^2-96\lambda^2+44g_2^2\lambda\xi+g_1^2(6g_2^2+4\lambda\xi)-g_2^4(\xi^2-9)]\\
k^2 \Sigma_{c_{HD}^2}^{\hat{{\mathcal Z}} \hat{\chi}}&-&i\frac{\sqrt{g_1^2+g_2^2}v_T}{2}\left[\Sigma^{\hat{\chi} \hat{\chi}}_{c_{HD}^2}+\frac{c_{HD}v_T^2}{4}\Sigma^{\hat{\chi} \hat{\chi}}_{c_{HD}}-\frac{c_{HD}^2v_T^4}{32}\Sigma^{\hat{\chi} \hat{\chi}}_{\rm SM}\right]+ i\frac{\sqrt{g_1^2+g_2^2}}{2}T^H_{c_{HD}^2}\, ,\nonumber\\
&=&c_{H\Box}c_{HD}\frac{iv_T^7}{1024\pi^2\epsilon}\sqrt{g_1^2+g_2^2}[96\lambda^2-44g_2^2\lambda\xi+g_2^4(\xi^2-9)+g_2^4(\xi^2-3)+2g_1^2(g_2^2[\xi^2-3]-18\lambda\xi)]\, .\nonumber\\
\end{eqnarray}
Again we do not expand every combination $c_ic_j$ here for the sake of brevity. The validity of the Ward identities has been confirmed for all combinations $c_ic_j$ using the amplitudes enumerated in Appendix~\ref{app:amps} and \cite{Corbett:2020ymv}. Table~\ref{tab:WIviolationsummary} summarizes which identities are violated for each combination $c_ic_j$ of Wilson coefficients when the field shifts are not properly handled. With proper care the Ward identities are preserved.

\section{Conclusions}\label{sec:conclusions}

In this work we present a new \FR package incorporating the full SMEFT Lagrangian at dimension-six and gauge fixed using the background field method. Additionally this package incorporates novel features not included in other publicly available \FR packages: dimension-eight operators of the forms of Class 2--4 as well as the dimension-six-squared effects of properly normalizing fields to $\mathcal{O}(1/\Lambda^4)$. As a test of the validity of the dimension-six part of the package a one-loop confirmation of the Ward identities was presented in a companion paper \cite{Corbett:2020ymv}. 

In extension of that work and to demonstrate the validity of the package to $\mathcal{O}(1/\Lambda^4)$ for operators of class 2--4 we have shown in this work the validity of the Ward identities at one loop to this order in the SMEFT power counting. Further, by defining the parameter $\Delta$ which accompanies the field shifts at dimension-six squared we were able to demonstrate the breakdown of the Ward identities when the field shifts are not properly implemented. The starkest example of the breakdown is the failure of $U(1)_{QED}$ in the presence of incomplete $c_{HW}^2$ corrections. We found that out of 21 combinations of Wilson coefficients, $c_ic_j$, one-third of them violate at least one of the Ward identities listed in Eqs.~\ref{eq:WIAA}--\ref{eq:WIZchi} when field renormalizations are not properly carried out. Violation of the Ward identities implies violation of gauge invariance, and, as the consistency of quantum field theories depends on gauge invariance, implies the unreliability of calculations. As many calculations performed in the literature include partial $\mathcal O(1/\Lambda^4)$ effects these calculations must be checked to ensure the consistency of analyses.

This package provides an analytic complement to the one-loop and tree-level simulation packages currently being developed \cite{Degrande:2020evl,Brivio:2017btx}. The appeal of the analytic approach is that step by step we may develop the full effective action at one-loop for the SMEFT with well defined cross checks such as those of the Ward identities demonstrated here. The appeal of the background field approach is the gauge invariant effective action, preservation of the Ward identities, and the various implications of this added symmetry as described in the introduction.

\section*{Acknowledgements}
TC acknowledges support from the Villum Fonden, project number 00010102,
 the Danish National Research Foundation through a DFF project grant, as well as funding from European Union's Horizon 2020 research and innovation programme under the Marie Sklodowska-Curie grant agreement No. 890787.
TC thanks A. Helset, M. Trott, and I. Brivio for useful discussions and their reading of the manuscript.

\newpage

%
%
%
%
%
%
%
%
%
%
%
\appendix
\section{Barred quantities}\label{app:bars}
The Feynrules files have built in substitutions which express barred quantities up to the corresponding power of \texttt{wilsonPower}.  The barred quantities in terms of unbarred quantities are derived using Eqs. 4.6-4.11 of \cite{Helset:2020yio}, and includes the following list of barred quantities:
\begin{equation}
\bar s_W,\ \ \ \ \bar c_W,\ \ \ \ \bar s_{\theta_Z},\ \ \ \ \bar c_{\theta_Z},\ \ \ \ \bar g_2\ \ \ \ \bar g_Z,\ \ \ \ \bar e,\ \ \ \ \bar M_Z,\ \ \ \ \bar M_W,\ \ \ \ \ \bar M_H\, .
\end{equation}
As it is relevant to Sec.~\ref{sec:ward} we reproduce the values of $\bar g_2$, $\bar g_Z$, $\bar M_W$, and $\bar M_Z$ here including the dimension-six squared dependence:
\begin{tiny}
\begin{eqnarray}
\bar g_2&=&g_2\left[1+c_{HW}v^2+\frac{3}{2}c_{HW}^2v^4\right]\, ,\\
\bar g_Z&=&\sqrt{g_1^2+g_2^2}\left[1+\frac{(g_1^2c_{HB}+g_1g_2c_{HWB}+g_2^2c_{HW})v^2}{g_1^2+g_2^2}
+\frac{(3g_1^4+4g_1^2g_2^2)c_{HB}^2+(4g_1^4+3g_2^4)c_{HW}+(g_1^4+g_1^2g_2^2+g_2^4)c_{HWB}^2}{2(g_1^2+g_2^2)^2}v^4\right.\nonumber\\
&&\left.\phantom{\sqrt{g_1^2+g_2^2}\hspace{4.1pt} 1}+\frac{g_1g_2[(g_1^2+2g_2^2)c_{HB}c_{HWB}+(2g_1^2+g_2^2)c_{HW}c_{HWB}-g_1g_2c_{HB}c_{HW}]}{(g_1^2+g_2^2)^2}v^4\right]\, ,\\
\bar M_W&=&\frac{g_2v}{2}\left[1+c_{HW}v^2+\frac{3}{2}c_{HW}^2v^2\right]\, ,\\
\frac{\bar M_Z}{\bar g_Z}&=&\frac{v}{2}\left[1+\frac{1}{4}c_{HD}v^2-\frac{1}{32}c_{HD}^2v^4\right]\, .
\end{eqnarray}
\end{tiny}

Upon loading the \FR package the substitution ``\texttt{barsubs}'' becomes available and can be used to change barred quantities into unbarred. It is automatically expanded to the preset value \texttt{wilsonPower}. Alternatively the substitution ``\texttt{tobars}'' may be used to rewrite $g_1$, $g_2$, and $g_3$ in terms of $\bar e$, $\bar g_2$, and $\bar g_3$. It is important to note that writing in terms of these barred quantities absorbs terms of higher order in the $\frac{1}{\Lambda^2}$ and therefore can convolute the form of the ward identities. The Ward identities are most easily verified in the unbarred basis of parameters.

%
%
%
%
%
%
%
%
%
%
%
\section{SMEFT operators used in this work}\label{app:ops}
This package includes the dimension-six SMEFT Lagrangian as reproduced in Table~\ref{tab:op59}, as well as the following dimension-eight operators organized according to the classes as defined at dimenions-six:
\begin{eqnarray}
\mathcal L_{\rm cl 2}&=&c_{H}^{(8)}(H^\dagger H)^4\, ,\\
\mathcal L_{\rm cl 3}&=&c_{HD}^{(8)}(H^\dagger H)^2(D_\mu H)^\dagger(D^\mu H)+c_{HD2}^{(8)}(H^\dagger H)(H^\dagger\sigma_a H)(D_\mu H)^\dagger\sigma^a(D^\mu H)\, ,\\
\mathcal L_{\rm cl 4}&=&c_{HB}^{(8)}(H^\dagger H)^2B^{\mu\nu}B_{\mu\nu}+c_{HW}(H^\dagger H)^2W_a^{\mu\nu}W^a_{\mu\nu}+(H^\dagger H)^2(H^\dagger\sigma^a H)(H^\dagger \sigma^b H)W_a^{\mu\nu}W_b^{\mu\nu}\nonumber\\
&&+c_{HWB}^{(8)}(H^\dagger H)^2(H^\dagger\sigma^a H)W_a^{\mu\nu}B_{\mu\nu}\, .
\end{eqnarray}

\begin{table}[H]
\begin{center}
\small
\begin{minipage}[t]{4.45cm}
\renewcommand{\arraystretch}{1.5}
\begin{tabular}[t]{c|c}
\multicolumn{2}{c}{$1:X^3$} \\
\hline
$Q_G$                & $f^{ABC} G_\mu^{A\nu} G_\nu^{B\rho} G_\rho^{C\mu} $ \\
$Q_W$                & $\epsilon^{IJK} W_\mu^{I\nu} W_\nu^{J\rho} W_\rho^{K\mu}$ \\
\end{tabular}
\end{minipage}
\begin{minipage}[t]{2.7cm}
\renewcommand{\arraystretch}{1.5}
\begin{tabular}[t]{c|c}
\multicolumn{2}{c}{$2:H^6$} \\
\hline
$Q_H$       & $(H^\dag H)^3$
\end{tabular}
\end{minipage}
\begin{minipage}[t]{5.1cm}
\renewcommand{\arraystretch}{1.5}
\begin{tabular}[t]{c|c}
\multicolumn{2}{c}{$3:H^4 D^2$} \\
\hline
$Q_{H\Box}$ & $(H^\dag H)\Box(H^\dag H)$ \\
$Q_{H D}$   & $\ \left(H^\dag D^\mu H\right)^* \left(H^\dag D_\mu H\right)$
\end{tabular}
\end{minipage}
\begin{minipage}[t]{2.7cm}

\renewcommand{\arraystretch}{1.5}
\begin{tabular}[t]{c|c}
\multicolumn{2}{c}{$4:X^2H^2$} \\
\hline
$Q_{H G}$     & $H^\dag H\, G^A_{\mu\nu} G^{A\mu\nu}$ \\
$Q_{H W}$     & $H^\dag H\, W^I_{\mu\nu} W^{I\mu\nu}$ \\
$Q_{H B}$     & $ H^\dag H\, B_{\mu\nu} B^{\mu\nu}$ \\
$Q_{H WB}$     & $ H^\dag \tau^I H\, W^I_{\mu\nu} B^{\mu\nu}$ \\
\end{tabular}
\end{minipage}

\vspace{0.15cm}

\begin{minipage}[t]{4.7cm}
\renewcommand{\arraystretch}{1.5}
\begin{tabular}[t]{c|c}
\multicolumn{2}{c}{$5: \psi^2H^3 + \hbox{h.c.}$} \\
\hline
$Q_{eH}$           & $(H^\dag H)(\bar l_p e_r H)$ \\
$Q_{uH}$          & $(H^\dag H)(\bar q_p u_r \widetilde H )$ \\
$Q_{dH}$           & $(H^\dag H)(\bar q_p d_r H)$\\
\end{tabular}
\end{minipage}
\begin{minipage}[t]{5.2cm}
\renewcommand{\arraystretch}{1.5}
\begin{tabular}[t]{c|c}
\multicolumn{2}{c}{$6:\psi^2 XH+\hbox{h.c.}$} \\
\hline
$Q_{eW}$      & $(\bar l_p \sigma^{\mu\nu} e_r) \tau^I H W_{\mu\nu}^I$ \\
$Q_{eB}$        & $(\bar l_p \sigma^{\mu\nu} e_r) H B_{\mu\nu}$ \\
$Q_{uG}$        & $(\bar q_p \sigma^{\mu\nu} T^A u_r) \widetilde H \, G_{\mu\nu}^A$ \\
$Q_{uW}$        & $(\bar q_p \sigma^{\mu\nu} u_r) \tau^I \widetilde H \, W_{\mu\nu}^I$ \\
$Q_{uB}$        & $(\bar q_p \sigma^{\mu\nu} u_r) \widetilde H \, B_{\mu\nu}$ \\
$Q_{dG}$        & $(\bar q_p \sigma^{\mu\nu} T^A d_r) H\, G_{\mu\nu}^A$ \\
$Q_{dW}$         & $(\bar q_p \sigma^{\mu\nu} d_r) \tau^I H\, W_{\mu\nu}^I$ \\
$Q_{dB}$        & $(\bar q_p \sigma^{\mu\nu} d_r) H\, B_{\mu\nu}$
\end{tabular}
\end{minipage}
\begin{minipage}[t]{5.4cm}
\renewcommand{\arraystretch}{1.5}
\begin{tabular}[t]{c|c}
\multicolumn{2}{c}{$7:\psi^2H^2 D$} \\
\hline
$Q_{H l}^{(1)}$      & $(H^\dag i\overleftrightarrow{D}_\mu H)(\bar l_p \gamma^\mu l_r)$\\
$Q_{H l}^{(3)}$      & $(H^\dag i\overleftrightarrow{D}^I_\mu H)(\bar l_p \tau^I \gamma^\mu l_r)$\\
$Q_{H e}$            & $(H^\dag i\overleftrightarrow{D}_\mu H)(\bar e_p \gamma^\mu e_r)$\\
$Q_{H q}^{(1)}$      & $(H^\dag i\overleftrightarrow{D}_\mu H)(\bar q_p \gamma^\mu q_r)$\\
$Q_{H q}^{(3)}$      & $(H^\dag i\overleftrightarrow{D}^I_\mu H)(\bar q_p \tau^I \gamma^\mu q_r)$\\
$Q_{H u}$            & $(H^\dag i\overleftrightarrow{D}_\mu H)(\bar u_p \gamma^\mu u_r)$\\
$Q_{H d}$            & $(H^\dag i\overleftrightarrow{D}_\mu H)(\bar d_p \gamma^\mu d_r)$\\
$Q_{H u d}$ + h.c.   & $i(\widetilde H ^\dag D_\mu H)(\bar u_p \gamma^\mu d_r)$\\
\end{tabular}
\end{minipage}

\vspace{0.15cm}

\begin{minipage}[t]{4.75cm}
\renewcommand{\arraystretch}{1.5}
\begin{tabular}[t]{c|c}
\multicolumn{2}{c}{$8:(\bar LL)(\bar LL)$} \\
\hline
$Q_{ll}$        & $(\bar l_p \gamma_\mu l_r)(\bar l_s \gamma^\mu l_t)$ \\
$Q_{qq}^{(1)}$  & $(\bar q_p \gamma_\mu q_r)(\bar q_s \gamma^\mu q_t)$ \\
$Q_{qq}^{(3)}$  & $(\bar q_p \gamma_\mu \tau^I q_r)(\bar q_s \gamma^\mu \tau^I q_t)$ \\
$Q_{lq}^{(1)}$                & $(\bar l_p \gamma_\mu l_r)(\bar q_s \gamma^\mu q_t)$ \\
$Q_{lq}^{(3)}$                & $(\bar l_p \gamma_\mu \tau^I l_r)(\bar q_s \gamma^\mu \tau^I q_t)$
\end{tabular}
\end{minipage}
\begin{minipage}[t]{5.25cm}
\renewcommand{\arraystretch}{1.5}
\begin{tabular}[t]{c|c}
\multicolumn{2}{c}{$8:(\bar RR)(\bar RR)$} \\
\hline
$Q_{ee}$               & $(\bar e_p \gamma_\mu e_r)(\bar e_s \gamma^\mu e_t)$ \\
$Q_{uu}$        & $(\bar u_p \gamma_\mu u_r)(\bar u_s \gamma^\mu u_t)$ \\
$Q_{dd}$        & $(\bar d_p \gamma_\mu d_r)(\bar d_s \gamma^\mu d_t)$ \\
$Q_{eu}$                      & $(\bar e_p \gamma_\mu e_r)(\bar u_s \gamma^\mu u_t)$ \\
$Q_{ed}$                      & $(\bar e_p \gamma_\mu e_r)(\bar d_s\gamma^\mu d_t)$ \\
$Q_{ud}^{(1)}$                & $(\bar u_p \gamma_\mu u_r)(\bar d_s \gamma^\mu d_t)$ \\
$Q_{ud}^{(8)}$                & $(\bar u_p \gamma_\mu T^A u_r)(\bar d_s \gamma^\mu T^A d_t)$ \\
\end{tabular}
\end{minipage}
\begin{minipage}[t]{4.75cm}
\renewcommand{\arraystretch}{1.5}
\begin{tabular}[t]{c|c}
\multicolumn{2}{c}{$8:(\bar LL)(\bar RR)$} \\
\hline
$Q_{le}$               & $(\bar l_p \gamma_\mu l_r)(\bar e_s \gamma^\mu e_t)$ \\
$Q_{lu}$               & $(\bar l_p \gamma_\mu l_r)(\bar u_s \gamma^\mu u_t)$ \\
$Q_{ld}$               & $(\bar l_p \gamma_\mu l_r)(\bar d_s \gamma^\mu d_t)$ \\
$Q_{qe}$               & $(\bar q_p \gamma_\mu q_r)(\bar e_s \gamma^\mu e_t)$ \\
$Q_{qu}^{(1)}$         & $(\bar q_p \gamma_\mu q_r)(\bar u_s \gamma^\mu u_t)$ \\
$Q_{qu}^{(8)}$         & $(\bar q_p \gamma_\mu T^A q_r)(\bar u_s \gamma^\mu T^A u_t)$ \\
$Q_{qd}^{(1)}$ & $(\bar q_p \gamma_\mu q_r)(\bar d_s \gamma^\mu d_t)$ \\
$Q_{qd}^{(8)}$ & $(\bar q_p \gamma_\mu T^A q_r)(\bar d_s \gamma^\mu T^A d_t)$\\
\end{tabular}
\end{minipage}

\vspace{0.15cm}

\begin{minipage}[t]{3.75cm}
\renewcommand{\arraystretch}{1.5}
\begin{tabular}[t]{c|c}
\multicolumn{2}{c}{$8:(\bar LR)(\bar RL)+\hbox{h.c.}$} \\
\hline
$Q_{ledq}$ & $(\bar l_p^j e_r)(\bar d_s q_{tj})$
\end{tabular}
\end{minipage}
\begin{minipage}[t]{5.5cm}
\renewcommand{\arraystretch}{1.5}
\begin{tabular}[t]{c|c}
\multicolumn{2}{c}{$8:(\bar LR)(\bar L R)+\hbox{h.c.}$} \\
\hline
$Q_{quqd}^{(1)}$ & $(\bar q_p^j u_r) \epsilon_{jk} (\bar q_s^k d_t)$ \\
$Q_{quqd}^{(8)}$ & $(\bar q_p^j T^A u_r) \epsilon_{jk} (\bar q_s^k T^A d_t)$ \\
$Q_{lequ}^{(1)}$ & $(\bar l_p^j e_r) \epsilon_{jk} (\bar q_s^k u_t)$ \\
$Q_{lequ}^{(3)}$ & $(\bar l_p^j \sigma_{\mu\nu} e_r) \epsilon_{jk} (\bar q_s^k \sigma^{\mu\nu} u_t)$
\end{tabular}
\end{minipage}
\begin{minipage}[t]{5.5cm}
\vskip 18mm
$\begin{aligned}
H^\dag i\overleftrightarrow{D}_\mu H  &\equiv H^\dag iD_\mu H - (i D_\mu H^\dag)H\\
H^\dag i\overleftrightarrow{D}^I_\mu H  &\equiv H^\dag i\tau^I D_\mu H - (i D_\mu \tau^I H^\dag)H
  \end{aligned}
$

\end{minipage}

\end{center}
\caption{\label{tab:op59}
The dimension-six operators of the SMEFT in the Warsaw basis, reproduced from \cite{Brivio:2017btx}. The CP-violating operators are not included as they are not currently implemented in this package. The operators are divided into eight Classes based on field content, those which are not hermitian are indiced by $+h.c.$ in the left-hand column. $p$, $r$, $s$, and $t$ are used as flavor indices for fermionic operators and are dropped in the operator labels $Q$.}
\end{table}

%
%
%
%
%
%
%
%
%
%
%
\newpage
\section{Dimension-six squared contributions to the amplitudes}\label{app:amps}
We enumerate the relevant amplitudes here. We consider only pairings of the Class 2-4 Wilson coefficients,
\begin{equation}\label{eq:quadlist}
c_H,\ \ \ \ \ c_{H\Box},\ \ \ \ \ c_{HD},\ \ \ \ \ c_{HB},\ \ \ \ \ c_{HW},\ \ \ \ \ c_{HWB}.
\end{equation}
In order to satisfy the Ward Identities, Eqs.~\ref{eq:wardids}, we need the following amplitudes:
\begin{equation}
\begin{array}{llll}
\Sigma_L^{\hat {\mathcal Z}\hat {\mathcal Z}}\, ,\ \ \ \ \ &\Sigma^{\hat{\mathcal  Z}\hat \chi}\, ,\ \ \ \ \ &\Sigma^{\hat\chi\hat\chi}\, ,\ \ \ \ \ \ \ \ \ \ &T^H\, ,\\
\vspace{.1cm}\Sigma_L^{\hat {\mathcal W}^+\hat {\mathcal W}^-}\, ,\ \ \ \ \ &\Sigma^{\hat \Phi^+\hat {\mathcal W}^-}\, ,\ \ \ \ \ &\Sigma^{\hat\Phi^+\hat\Phi^-}\, .
\end{array}
\end{equation}
All amplitudes are calculated using the \FR package presented here, fed into \FA \cite{Hahn:2000kx} and \FC \cite{Hahn:1998yk}, with UV divergent parts being calculated using Package-X \cite{Patel:2016fam} (which automates dimensional regularization). Details of streamlining this workflow are included in Appendix~\ref{app:speedingup} as well as the ancillary files. In the case that shifts in field normalization at dimension-six squared are involved in the calculation, they are flagged with $\Delta$ to indicate their contribution. In Section~\ref{sec:verifyWI} these $\Delta$s are used to demonstrate the necessity of these contributions to maintain the Ward identities. The full quadratic contribution to the amplitude can be obtained by simply taking $\Delta\to1$. We do not include $\Sigma_T$ here for brevity, the corresponding expressions quickly become unmanageable for two insertions of the effective operators and are not relevant to the discussion on the Ward identities. Also, we only include the $\Delta$ dependence for the $c_i^2$ contributions, for $c_ic_j$ with $i\ne j$ we set $\Delta\to1$.

\subsection*{$c_H^2$ contributions}
\begin{equation}
\begin{array}{rcl}
\Sigma^{\hat{\mathcal A}\hat{\mathcal A}}&=&0
=\Sigma^{\hat{\mathcal A}\hat{\mathcal Z}}
=\Sigma^{\hat{\mathcal Z}\hat{\mathcal Z}}
=\Sigma^{\hat{\mathcal W^+}\hat{\mathcal W^-}}
=\Sigma^{\hat{\mathcal Z}\hat \chi}\\[16pt]
\Sigma^{\hat\chi\hat\chi}&=&c_H^2\frac{45v_T^6}{32\pi^2\epsilon}\\[16pt]
\Sigma^{\hat{\phi}^+\hat{\phi}^-}&=&c_H^2\frac{45v_T^6}{32\pi^2\epsilon}\\[16pt]
T^H&=&c_H^2\frac{45v_T^6}{32\pi^2\epsilon}
\end{array}
\end{equation}

\subsection*{$c_{H\Box}^2$ contributions}
\begin{equation}
\begin{array}{rcl}
\Sigma^{\hat{\mathcal A}\hat{\mathcal A}}&=&0=\Sigma^{\hat{\mathcal A}\hat{\mathcal Z}}\\[16pt]
\Sigma^{\hat{\mathcal Z}\hat{\mathcal Z}}_L&=&c_{H\Box}^2\frac{3(g_1^2+g_2^2)^2v_T^6}{256\pi^2\epsilon}(5-\Delta)\\[16pt]
\Sigma^{\hat{\mathcal W^+}\hat{\mathcal W^-}}_L&=&c_{H\Box}^2\frac{9v_T^6g_2^4}{768\pi^2\epsilon}(5-\Delta)\\[16pt]
\Sigma^{\hat{\mathcal Z}\hat \chi}&=&c_{H\Box}^2\frac{iv_T^5(g_1^2+g_2^2)^{\frac{3}{2}}}{128\pi^2\epsilon}[-(15+\xi)+(3+\xi)\Delta]\\[16pt]
\Sigma^{\hat{\phi}^+ \hat{\mathcal{W}}^-}&=&c_{H\Box}^2\frac{v_T^5g_2^3}{128\pi^2\epsilon}[-(15+\xi)+(3+\xi)]\\[16pt]
\Sigma^{\hat\chi\hat\chi}&=&c_{H\Box}^2\frac{v_T^4}{256\pi^2\epsilon}\Big[v_T^2(1920\lambda^2-32[g_1^2+g_2^2]\lambda\xi+[g_1^2+g_2^2]^2\xi^2)-4(g_1^2+g_2^2)(15+\xi)k^2\\
&&\phantom{c_{H\Box}^2\frac{v_T^4}{256\pi^2\epsilon}\Big[}-(g_1^2+g_2^2)v_T^2\xi(-32\lambda+[g_1^2+g_2^2]\xi)\Delta+4(g_1^2+g_2^2)(3+\xi)k^2\Delta\Big]\\[16pt]
\Sigma^{\hat{\phi}^+\hat{\phi}^-}&=&c_{H\Box}^2\frac{v_T^4}{256\pi^2\epsilon}\Big[v_T^2(1920\lambda^2-32g_2^2\lambda\xi+g_2^4\xi^2)-4g_2^2(15+\xi)k^2\\
&&\phantom{c_{H\Box}^2\frac{v_T^4}{256\pi^2\epsilon}\Big[}-g_2^2(v_T^2\xi[-32\lambda+g_2^2\xi]-4[3+\xi]k^2)\Delta\Big]\\[16pt]
T^H&=&c_{H\Box}^2\frac{v_T^7}{512\pi^2\epsilon}\Big[4(3g_1^4+6g_1^2g_2^2+9g_2^4+1216\lambda^2+4[g_1^2+3g_2^2]\lambda\xi)\\
&&\phantom{c_{H\Box}^2\frac{v_T^7}{512\pi^2\epsilon}\Big[}+(-3[g_1^4+2g_1^2g_2^2+3g_2^4-96\lambda^2]-4[g_1^2+3g_2^2]\lambda\xi)\Delta\Big]
\end{array}
\end{equation}

\subsection*{$c_{HD}^2$ contributions}
\begin{equation}
\begin{array}{rcl}
\Sigma^{\hat{\mathcal A}\hat{\mathcal A}}&=&0=\Sigma^{\hat{\mathcal A}\hat{\mathcal Z}}\\[16pt]
\Sigma^{\hat{\mathcal Z}\hat{\mathcal Z}}_L&=&c_{HD}^2\frac{v_T^4(g_1^2+g_2^2)}{4096\pi^2\epsilon}\Big[([15g_1^2+39g_2^2-232\lambda]v_T^2-16k^2)+(-3[g_1^2+g_2^2]+8\lambda)v_T^2\Delta\Big]\\[16pt]
\Sigma^{\hat{\mathcal W^+}\hat{\mathcal W^-}}_L&=&c_{HD}^2\frac{v_T^4g_2^2}{2048\pi^2\epsilon}\Big[(3[2g_1^2+5g_2^2]v_T^2-8k^2)-3g_2^2v_T^2\Delta\Big]\\[16pt]
\Sigma^{\hat{\mathcal Z}\hat \chi}&=&c_{HD}^2\frac{i v_T^3\sqrt{g_1^2+g_2^2}}{4096\pi^2\epsilon}\Big[(2v_T^2[272\lambda+g_1^2(\xi-3)+5g_2^2(\xi-3)]+32k^2)+(3g_1^2+5g_2^2)v_T^2(3+\xi)\Delta\Big]\\[16pt]
\Sigma^{\hat{\phi}^+ \hat{\mathcal{W}}^-}&=&c_{HD}^2\frac{v_T^3g_2}{1024\pi^2\epsilon}\Big[(-v_T^2[6g_1^2+g_2^2(15+\xi)]+8k^2)+g_2^2v_T^2(3+\xi)\Delta\Big]\\[16pt]
\Sigma^{\hat\chi\hat\chi}&=&c_{HD}^2\frac{v_T^2}{4096\pi^2\epsilon}\Big[v_T^4(3488\lambda^2+36g_2^2\lambda\xi+g_1^4[6+\xi^2]+g_2^4[36+\xi^2]+2g_1^2[-2\lambda\xi+g_2^2(6+\xi^2)])\\
&&\phantom{c_{HD}^2\frac{v_T^2}{4096\pi^2\epsilon}\Big[}+8k^2v_T^2(160\lambda+4g_2^2\xi+g_1^2[3+\xi])+64k^4\\
&&\phantom{c_{HD}^2\frac{v_T^2}{4096\pi^2\epsilon}\Big[}-v_T^4[6(g_1^4+2g_1^2g_2^2+2g_2^4)+160\lambda^2-4(5g_1^2+3g_2^2)\lambda\xi+(g_1^2+g_2^2)^2\xi^2]\Delta\\
&&\phantom{c_{HD}^2\frac{v_T^2}{4096\pi^2\epsilon}\Big[}+8[g_1^2+2g_2^2]v_T^2[3+\xi]k^2\Delta\Big]\\[16pt]
\Sigma^{\hat{\phi}^+\hat{\phi}^-}&=&c_{HD}^2\frac{v_T^2}{4096\pi^2\epsilon}\Big[v_T^4(27g_1^4+1920\lambda^2-24g_2^2\lambda\xi+g_2^4[27+2\xi^2]+g_1^2[8\lambda\xi+g_2^2(42+4\xi^2)])\\
&&\phantom{c_{HD}^2\frac{v_T^2}{4096\pi^2\epsilon}\Big[}-8v_T^2(6g_1^2+g_2^2[15+\xi])k^2+64k^4\\
&&\phantom{c_{HD}^2\frac{v_T^2}{4096\pi^2\epsilon}\Big[}-v_T^4(3[g_1^2+g_2^2]^2+8[g_1^2-3g_2^2]\lambda\xi-4g_1^2g_2^2[3-\xi^2]+2g_2^4\xi^2)\Delta+8g_2^2v_T^2(3+\xi)k^2\Delta\Big]\\[16pt]
T^H&=&c_{HD}^2\frac{v_T^7}{8192\pi^2\epsilon}\Big[2(2432\lambda^2+32g_2^2\lambda\xi+g_1^4[15+\xi^2]+g_2^4[27+\xi^2]+2g_1^2[8\lambda\xi+g_2^2(15+\xi^2)])\\
&&\phantom{c_{HD}^2\frac{v_T^7}{8192\pi^2\epsilon}\Big[}+(-3[3g_1^4+6g_1^2g_2^2+5g_2^4-96\lambda^2]-4[5g_1^2+7g_2^2]\lambda\xi-2[g_1^2+g_2^2]^2\xi^2)\Delta\Big]
\end{array}
\end{equation}

\newpage
\subsection*{$c_{HB}^2$ contributions}
\begin{equation}\label{eq:cHB2}
\begin{array}{rcl}
\Sigma^{\hat{\mathcal A}\hat{\mathcal A}}_L&=&0\\[16pt]
\Sigma^{\hat{\mathcal A}\hat{\mathcal Z}}_L&=&-c_{HB}^2\frac{g_1g_2(g_1^2+3g_2^2)v_T^6(3+\xi)}{128\pi^2\epsilon}\left[1-
\Delta\right]\\[16pt]
\Sigma^{\hat{\mathcal Z}\hat{\mathcal Z}}_L&=&c_{HB}^2\frac{g_1^2v_T^6}{128\pi^2\epsilon}\left[(7g_1^2+10g_2^2)(3+\xi)-(g_1^2+2g_2^2)(3+\xi)\Delta\right]\\[16pt]
\Sigma^{\hat{\mathcal W^+}\hat{\mathcal W^-}}_L&=&c_{HB}^2\frac{g_1^2g_2^2v_T^6}{256\pi^2\epsilon}\left[5(3+\xi)-(3+\xi)\Delta\right]\\[16pt]
\Sigma^{\hat{\mathcal Z}\hat \chi}&=&c_{HB}^2\frac{i v_T^5}{256\pi^2\epsilon(g_1^2+g_2^2)^{\frac{3}{2}}}\left[-(18g_1^6+41g_1^4g_2^2+25g_1^2g_2^4)(3+\xi)+g_1^2(g_1^2+g_2^2)(3g_1^2+5g_2^2)(3+\xi)\Delta\right]\\[16pt]
\Sigma^{\hat{\phi}^+ \hat{\mathcal{W}}^-}&=&c_{HB}^2\frac{v_T^5g_1^2g_2}{128\pi^2\epsilon}\left[-5(3+\xi)+(3+\xi)\Delta\right]\\[16pt]
\Sigma^{\hat\chi\hat\chi}&=&c_{HB}^2\frac{v_T^4g_1^2}{128\pi^2\epsilon}\Big[v_T^2(-10\lambda\xi+g_2^2[21+\xi^2]+g_1^2[33+\xi^2])-10(3+\xi)k^2\\
&&\phantom{c_{HB}^2\frac{v_T^4g_1^2}{128\pi^2\epsilon}\Big[}+(v_T^2[3(g_1^2+g_2^2)+18\lambda\xi-(g_1^2+g_2^2)\xi^2]+2[3+\xi]k^2)\Delta\Big]\\[16pt]
\Sigma^{\hat{\phi}^+\hat{\phi}^-}&=&c_{HB}^2\frac{v_T^4g_1^2}{256\pi^2\epsilon}\Big[v_T^2(4\lambda\xi+g_2^2[66-9\xi^2]+2g_1^2[33+\xi^2])-20(3+\xi)k^2\\
&&\phantom{c_{HB}^2\frac{v_T^4g_1^2}{256\pi^2\epsilon}\Big[}+(v_T^2[12\lambda\xi-2g_1^2(-3+\xi^2)+9g_2^2(-2+\xi^2)]+4[3+\xi]k^2)\Delta\Big]\\[16pt]
T^H&=&c_{HB}^2\frac{v_T^7g_1^2}{256\pi^2\epsilon}\left[66g_1^2+42g_2^2+4\lambda\xi-(g_1^2+g_2^2)\xi^2+(12\lambda\xi+g_1^2[6+\xi^2]+g_2^2[6+\xi^2])\Delta\right]
\end{array}
\end{equation}

\newpage
\subsection*{$c_{HW}^2$ contributions}\label{eq:cHW2}
\begin{equation}
\begin{array}{rcl}
\Sigma^{\hat{\mathcal A}\hat{\mathcal A}}_L&=&-c_{HW}^2\frac{g_1^2g_2^4v_T^6}{128\pi^2\epsilon(g_1^2+g_2^2)\xi}(3+9\xi-7\xi^2+3\xi^3)(1-\Delta)\\[16pt]
\Sigma^{\hat{\mathcal A}\hat{\mathcal Z}}_L&=&c_{HW}^2\frac{g_1g_2v_T^6}{128\pi^2\epsilon(g_1^2+g_2^2)\xi}\Big[g_1^4\xi(3+\xi)+4g_1^2g_2^2\xi(3+\xi)+g_2^4(-3+[10-3\xi]\xi^2\Big](1-\Delta)\\[16pt]
\Sigma^{\hat{\mathcal Z}\hat{\mathcal Z}}_L&=&c_{HW}^2\frac{g_2^2v_T^6}{128\pi^2\epsilon(g_1^2+g_2^2)\xi}\Big[10g_1^4\xi(3+\xi)+31g_1^2g_2^2\xi(3+\xi)+g_2^4(-3+54\xi+28\xi^2-3\xi^3)\\
&&\phantom{c_{HW}^2\frac{g_2^2v_T^6}{128\pi^2\epsilon(g_1^2+g_2^2)\xi}\Big[}-(2g_1^4\xi[3+\xi]+5g_1^2g_2^2\xi[3+\xi]+g_2^4[-3+10\xi^2-3\xi^3])\Delta\Big]\\[16pt]
\Sigma^{\hat{\mathcal W^+}\hat{\mathcal W^-}}_L&=&c_{HW}^2\frac{g_2^2v_T^6}{256\pi^2\epsilon\xi}\Big[5g_1^2\xi(3+\xi)+2g_2^2(-3+54\xi+28\xi^2-3\xi^3)+(-g_1^2\xi[3+\xi]+g_2^2[6-20\xi^2+6\xi^3])\Delta\Big]\\[16pt]
\Sigma^{\hat{\mathcal Z}\hat \chi}&=&c_{HW}^2\frac{ig_2^2v_T^5}{256\pi^2\epsilon(g_1^2+g_2^2)^\frac{3}{2}\xi}\Big[-(35g_1^4+91g_1^2g_2^2+54g_2^4)\xi(3+\xi)+(g_1^2+g_2^2)(7g_1^2+9g_2^2)\xi(3+\xi)\Delta\Big]\\[16pt]
\Sigma^{\hat{\phi}^+ \hat{\mathcal{W}}^-}&=&-c_{HW}^2\frac{v_T^5}{256\pi^2\epsilon}\Big[2(2g_1^2g_2+27g_2^3)(3+\xi)-g_2(g_1^2+9g_2^2)(3+\xi)\Delta\Big]\\[16pt]
\Sigma^{\hat\chi\hat\chi}&=&c_{HW}^2\frac{g_2^2v_T^4}{128\pi^2\epsilon}\Big[v_T^2(99g_2^2-6\lambda\xi+g_1^2[21+\xi^2])-30(3+\xi)k^2\\
&&\phantom{c_{HW}^2\frac{g_2^2v_T^6}{256\pi^2\epsilon}\Big[}+(v_T^2[9g_2^2+30\lambda\xi-g_1^2(-3+\xi^2)]+6[3+\xi]k^2)\Delta\Big]\\[16pt]
\Sigma^{\hat{\phi}^+\hat{\phi}^-}&=&3c_{HW}^2\frac{g_2^2v_T^4}{256\pi^2\epsilon}\Big[v_T^2(66g_2^2-4\lambda\xi+g_1^2[22-3\xi^2])-20(3+\xi)k^2\\
&&\phantom{3c_{HW}^2\frac{g_2^2v_T^4}{256\pi^2\epsilon}\Big[}+(v_T^2[6g_2^2+20\lambda\xi+3g_1^2(-2+\xi^2)]+4[3+\xi]k^2)\Delta\Big]\\[16pt]
T^H&=&c_{HW}^2\frac{g_2^2v_T^7}{256\pi^2\epsilon}\Big[12\lambda\xi-3g_2^2(-66+\xi^2)-g_1^2(-42+\xi^2)+(36\lambda\xi+g_1^2[6+\xi^2]+3g_2^2[6+\xi^2])\Delta\Big]
\end{array}
\end{equation}

\newpage
\subsection*{$c_{HWB}^2$ contributions}
\begin{equation}
\begin{array}{rcl}
\Sigma^{\hat{\mathcal A}\hat{\mathcal A}}_L&=&0=\Sigma^{\hat{\mathcal A}\hat{\mathcal Z}}_L\\[16pt]
\Sigma^{\hat{\mathcal Z}\hat{\mathcal Z}}_L&=&c_{HWB}^2\frac{v_T^6}{512\pi^2\epsilon}\Big[(5g_1^4+23g_1^2g_2^2+10g_2^4)(3+\xi)-(g_1^2+g_2^2)(g_1^2+2g_2^2)(3+\xi)\Delta\Big]\\[16pt]
\Sigma^{\hat{\mathcal W^+}\hat{\mathcal W^-}}_L&=&c_{HWB}^2\frac{v_T^4g_2^2}{1024\pi^2\epsilon\xi}\Big[v_T^2(g_1^2\xi[39+5\xi]-3g_2^2[1-10\xi-4\xi^2+\xi^3])-32\xi k^2\\
&&\phantom{c_{HWB}^2\frac{v_T^4g_2^2}{1024\pi^2\epsilon}\Big[}+v_T^2(-g_1^2\xi[3+\xi]+g_2^2[3+6\xi-8\xi^2+3\xi^3])\Delta\Big]\\[16pt]
\Sigma^{\hat{\mathcal Z}\hat \chi}&=&c_{HWB}^2\frac{iv_T^5(3+\xi)}{1024\pi^2\epsilon(g_1^2+g_2^2)^\frac{3}{2}}\Big[-15g_1^6-67g_1^4g_2^2-69g_1^2g_2^4-25g_2^6+(g_1^2+g_2^2)^2(3g_1^2+5g_2^2)\Delta\Big]\\[16pt]
\Sigma^{\hat{\phi}^+ \hat{\mathcal{W}}^-}&=&c_{HWB}^2\frac{g_2v_T^3}{512\pi^2\epsilon}\Big[-(g_1^2+g_2^2)v_T^2(39+5\xi)+32k^2+(g_1^2+g_2^2)v_T^2(3+\xi)\Delta\Big]\\[16pt]
\Sigma^{\hat\chi\hat\chi}&=&c_{HWB}^2\frac{v_T^4}{512\pi^2\epsilon}\Big[v_T^2(3[7g_1^4+30g_1^2g_2^2+7g_2^4]-10[g_1^2+g_2^2]\lambda\xi+[g_1^2+g_2^2]^2\xi^2)-10(g_1^2+g_2^2)(3+\xi)k^2\\
&&\phantom{c_{HWB}^2\frac{v_T^4}{512\pi^2\epsilon}\Big[}-(g_1^2+g_2^2)(v_T^2[-18\lambda\xi+g_1^2(\xi^2-3)+g_2^2(\xi^2-3)]-2[3+\xi]k^2)\Delta\Big]\\[16pt]
\Sigma^{\hat{\phi}^+\hat{\phi}^-}&=&c_{HWB}^2\frac{v_T^2}{1024\pi^2\epsilon}\Big[v_T^4(4g_2^2\lambda\xi-g_2^4[-42+\xi^2]+2g_1^4[21+\xi^2]+g_1^2[4\lambda\xi+g_2^2(216-11\xi^2)])\\
&&\phantom{c_{HWB}^2\frac{v_T^2}{1024\pi^2\epsilon}\Big[}-4(g_1^2+g_2^2)v_T^2(39+5\xi)k^2+128k^4\\
&&\phantom{c_{HWB}^2\frac{v_T^2}{1024\pi^2\epsilon}\Big[}+v_T^4(6[g_1^4-4g_1^2g_2^2+g_2^4]+12[g_1^2+g_2^2]\lambda\xi+[-2g_1^4+11g_1^2g_2^2+g_2^4]\xi^2\Delta\\
&&\phantom{c_{HWB}^2\frac{v_T^2}{1024\pi^2\epsilon}\Big[}+4(g_1^2+g_2^2)v_T^2(3+\xi)k^2)\Delta\Big]\\[16pt]
T^H&=&c_{HWB}^2\frac{v_T^7}{1024\pi^2\epsilon}\Big[6(7g_1^4+30g_1^2g_2^2+7g_2^4)+4(g_1^2+g_2^2)\lambda\xi-(g_1^2+g_2^2)^2\xi^2\\
&&\phantom{c_{HWB}^2\frac{v_T^7}{1024\pi^2\epsilon}\Big[}+(g_1^2+g_2^2)(12\lambda\xi+g_1^2[6+\xi^2]+g_2^2[6+\xi^2])\Delta\Big]
\end{array}
\end{equation}

\vfill
\subsection*{$c_Hc_{H\Box}$ contributions}
Next we consider the mixed terms, beginning with $c_H c_{H\Box}$. For compactness we take $\Delta\to1$ in the remaining amplitudes:
\begin{equation}
\begin{array}{rcl}
\Sigma^{\hat{\mathcal A}\hat{\mathcal A}}&=&0=\Sigma^{\hat{\mathcal A}\hat{\mathcal Z}}=
\Sigma^{\hat{\mathcal Z}\hat{\mathcal Z}}=
\Sigma^{\hat{\mathcal W^+}\hat{\mathcal W^-}}=
\Sigma^{\hat{\mathcal Z}\hat \chi}=
\Sigma^{\hat{\phi}^+ \hat{\mathcal{W}}^-}\\[16pt]
\Sigma^{\hat\chi\hat\chi}&=&-c_Hc_{H\Box}\frac{15 \lambda v_T^6}{2\pi^2\epsilon}\\[16pt]
\Sigma^{\hat{\phi}^+\hat{\phi}^-}&=&-c_Hc_{H\Box}\frac{15 \lambda v_T^6}{2\pi^2\epsilon}\\[16pt]
T^H&=&-c_Hc_{H\Box}\frac{3v_T^7}{128\pi^2\epsilon}[384\lambda+(g_1^2+3g_2^2)\xi]\\
\end{array}
\end{equation}

\newpage
\subsection*{$c_Hc_{HD}$ contributions}
\begin{equation}
\begin{array}{rcl}
\Sigma^{\hat{\mathcal A}\hat{\mathcal A}}&=&0=\Sigma^{\hat{\mathcal A}\hat{\mathcal Z}}=\Sigma^{\hat{\mathcal W^+}\hat{\mathcal W^-}}=\Sigma^{\hat{\phi}^+ \hat{\mathcal{W}}^-}\\[16pt]
\Sigma^{\hat{\mathcal Z}\hat{\mathcal Z}}_T&=&\Sigma^{\hat{\mathcal Z}\hat{\mathcal Z}}_L=-c_Hc_{HD}\frac{9v_T^6}{128\pi^2\epsilon}(g_1^2+g_2^2)\\[16pt]
\Sigma^{\hat{\mathcal Z}\hat \chi}&=&c_Hc_{HD}\frac{i9v_T^5}{64\pi^2\epsilon}\sqrt{g_1^2+g_2^2}\\[16pt]
\Sigma^{\hat\chi\hat\chi}&=&c_Hc_{HD}\frac{3v_T^4}{256\pi^2\epsilon}[v_T^2(224\lambda+[g_1^2+3g_2^2]\xi)+24k^2]\\[16pt]
\Sigma^{\hat{\phi}^+\hat{\phi}^-}&=&c_Hc_{HD}\frac{15\lambda v_T^6}{8\pi^2\epsilon}\\[16pt]
T^H&=&c_Hc_{HD}\frac{3v_T^7}{512\pi^2\epsilon}[384\lambda+(g_1^2+3g_2^2)\xi]
\end{array}
\end{equation}
\vfill

\subsection*{$c_Hc_{HB}$ contributions}
\begin{equation}
\begin{array}{rcl}
\Sigma^{\hat{\mathcal A}\hat{\mathcal A}}_L&=&0=\Sigma^{\hat{\mathcal A}\hat{\mathcal Z}}_L=\Sigma^{\hat{\mathcal W^+}\hat{\mathcal W^-}}_L=\Sigma^{\hat{\mathcal Z}\hat{\mathcal Z}}_L=\Sigma^{\hat{\mathcal Z}\hat \chi}=\Sigma^{\hat{\phi}^+ \hat{\mathcal{W}}^-}\\[16pt]
\Sigma^{\hat\chi\hat\chi}&=&-c_Hc_{HB}\frac{3g_1^2v_T^6\xi}{64\pi^2\epsilon}\\[16pt]
\Sigma^{\hat{\phi}^+\hat{\phi}^-}&=&-c_Hc_{HB}\frac{3g_1^2v_T^6\xi}{64\pi^2\epsilon}\\[16pt]
T^H&=&-c_Hc_{HB}\frac{3g_1^2v_T^7\xi}{64\pi^2\epsilon}
\end{array}
\end{equation}
\vfill

\subsection*{$c_{H}c_{HW}$ contributions}
\begin{equation}
\begin{array}{rcl}
\Sigma^{\hat{\mathcal A}\hat{\mathcal A}}_L&=&\Sigma^{\hat{\mathcal A}\hat{\mathcal Z}}_L=\Sigma^{\hat{\mathcal Z}\hat{\mathcal Z}}_L=
\Sigma^{\hat{\mathcal W^+}\hat{\mathcal W^-}}_L=\Sigma^{\hat{\mathcal Z}\hat \chi}=\Sigma^{\hat{\phi}^+ \hat{\mathcal{W}}^-}\\[16pt]
\Sigma^{\hat\chi\hat\chi}&=&-c_Hc_{HW}\frac{9g_2^2v_T^6\xi}{64\pi^2\epsilon}\\[16pt]
\Sigma^{\hat{\phi}^+\hat{\phi}^-}&=&-c_Hc_{HW}\frac{9g_2^2v_T^6\xi}{64\pi^2\epsilon}\\[16pt]
T^H&=&-c_Hc_{HW}\frac{9g_2^2v_T^7\xi}{64\pi^2\epsilon}
\end{array}
\end{equation}

\newpage
\subsection*{$c_{H}c_{HWB}$ contributions}
\begin{equation}
\begin{array}{rcl}
\Sigma^{\hat{\mathcal A}\hat{\mathcal A}}_L&=&\Sigma^{\hat{\mathcal A}\hat{\mathcal Z}}_L=\Sigma^{\hat{\mathcal Z}\hat{\mathcal Z}}_L=
\Sigma^{\hat{\mathcal W^+}\hat{\mathcal W^-}}_L=\Sigma^{\hat{\mathcal Z}\hat \chi}=\Sigma^{\hat{\phi}^+ \hat{\mathcal{W}}^-}\\[16pt]
\Sigma^{\hat\chi\hat\chi}&=&-c_Hc_{HWB}\frac{3g_1g_2v_T^6\xi}{64\pi^2\epsilon}\\[16pt]
\Sigma^{\hat{\phi}^+\hat{\phi}^-}&=&-c_Hc_{HWB}\frac{3g_1g_2v_T^6\xi}{64\pi^2\epsilon}\\[16pt]
T^H&=&-c_Hc_{HWB}\frac{3g_1g_2v_T^7\xi}{64\pi^2\epsilon}
\end{array}
\end{equation}
\vfill

\subsection*{$c_{H\Box}c_{HD}$ contributions}
\begin{equation}\label{eq:cHBoxcHD}
\begin{array}{rcl}
\Sigma^{\hat{\mathcal A}\hat{\mathcal A}}_L&=&0=\Sigma^{\hat{\mathcal A}\hat{\mathcal Z}}_L\\[16pt]
\Sigma^{\hat{\mathcal Z}\hat{\mathcal Z}}_L&=&c_{H\Box}c_{HD}\frac{3v_T^6}{128\pi^2\epsilon}(g_1^2+g_2^2)(g_1^2+g_2^2+8\lambda)\\[16pt]
\Sigma^{\hat{\mathcal W^+}\hat{\mathcal W^-}}_L&=&-c_{H\Box}c_{HD}\frac{3g_2^4v_T^6}{128\pi^2\epsilon}\\[16pt]
\Sigma^{\hat{\mathcal Z}\hat \chi}&=&-c_{H\Box}c_{HD}\frac{3iv_T^5}{256\pi^2\epsilon}\sqrt{g_1^2+g_2^2}(3g_1^2+3g_2^2+32\lambda)\\[16pt]
\Sigma^{\hat{\phi}^+ \hat{\mathcal{W}}^-}&=&c_{H\Box}c_{HD}\frac{3g_2^3v_T^5}{64\pi^2\epsilon}\\[16pt]
\Sigma^{\hat\chi\hat\chi}&=&-c_{H\Box}c_{HD}\frac{v_T^4}{64\pi^2\epsilon}[304\lambda^2v_T^2+3(g_1^2+g_2^2+16\lambda)k^2]\\[16pt]
\Sigma^{\hat{\phi}^+\hat{\phi}^-}&=&c_{H\Box}c_{HD}\frac{3v_T^4}{32\pi^2\epsilon}(-40\lambda^2v_T^2+g_2^2k^2)\\[16pt]
T^H&=&c_{H\Box}c_{HD}\frac{v_T^7}{1024\pi^2\epsilon}[9g_1^4-9g_2^4-5152\lambda^2-36g_2^2\lambda\xi+6g_1^2(3g_2^2-2\lambda\xi)]
\end{array}
\end{equation}
\vfill

\subsection*{$c_{H\Box}c_{HB}$ contributions}
\begin{equation}
\begin{array}{rcl}
\Sigma^{\hat{\mathcal A}\hat{\mathcal A}}_L&=&0=\Sigma^{\hat{\mathcal A}\hat{\mathcal Z}}_L=\Sigma^{\hat{\mathcal W^+}\hat{\mathcal W^-}}_L=\Sigma^{\hat{\phi}^+ \hat{\mathcal{W}}^-}=\Sigma^{\hat{\phi}^+\hat{\phi}^-}\\[16pt]
\Sigma^{\hat{\mathcal Z}\hat{\mathcal Z}}_L&=&c_{H\Box}c_{HB}\frac{3g_1^2v_T^6}{32\pi^2\epsilon}(g_1^2+g_2^2)\\[16pt]
\Sigma^{\hat{\mathcal Z}\hat \chi}&=&-c_{H\Box}c_{HB}\frac{9ig_1^2v_T^5}{64\pi^2\epsilon}\sqrt{g_1^2+g_2^2}\\[16pt]
\Sigma^{\hat\chi\hat\chi}&=&-c_{H\Box}c_{HB}\frac{3g_1^2v_T^4k^2}{16\pi^2\epsilon}\\[16pt]
T^H&=&c_{H\Box}c_{HB}\frac{g_1^2v_T^7}{128\pi^2\epsilon}(9g_1^2+9g_2^2+4\lambda\xi)
\end{array}
\end{equation}

\newpage
\subsection*{$c_{H\Box}c_{HW}$ contributions}
\begin{equation}
\begin{array}{rcl}
\Sigma^{\hat{\mathcal A}\hat{\mathcal A}}_L&=&0=\Sigma^{\hat{\mathcal A}\hat{\mathcal Z}}_L\\[16pt]
\Sigma^{\hat{\mathcal Z}\hat{\mathcal Z}}_L&=&c_{H\Box}c_{HW}\frac{3g_2^2v_T^6}{32\pi^2\epsilon}(g_1^2+g_2^2)\\[16pt]
\Sigma^{\hat{\mathcal W^+}\hat{\mathcal W^-}}_L&=&c_{H\Box}c_{HW}\frac{3g_2^4v_T^6}{32\pi^2\epsilon}\\[16pt]
\Sigma^{\hat{\mathcal Z}\hat \chi}&=&-c_{H\Box}c_{HW}\frac{9ig_2^2v_T^5}{64\pi^2\epsilon}\sqrt{g_1^2+g_2^2}\\[16pt]
\Sigma^{\hat{\phi}^+ \hat{\mathcal{W}}^-}&=&-c_{H\Box}c_{HW}\frac{9g_2^3v_T^5}{64\pi^2\epsilon}\\[16pt]
\Sigma^{\hat\chi\hat\chi}&=&-c_{H\Box}c_{HW}\frac{3g_2^2v_T^4k^2}{16\pi^2\epsilon}\\[16pt]
\Sigma^{\hat{\phi}^+\hat{\phi}^-}&=&-c_{H\Box}c_{HW}\frac{3g_2^2v_T^4k^2}{16\pi^2\epsilon}\\[16pt]
T^H&=&c_{H\Box}c_{HW}\frac{3g_2^2v_T^7}{128\pi^2\epsilon}(3g_1^2+9g_2^2+4\lambda\xi)
\end{array}
\end{equation}

\vfill
\subsection*{$c_{H\Box}c_{HWB}$ contributions}
\begin{equation}
\begin{array}{rcl}
\Sigma^{\hat{\mathcal A}\hat{\mathcal A}}_L&=&0=\Sigma^{\hat{\mathcal A}\hat{\mathcal Z}}_L=\Sigma^{\hat{\mathcal W^+}\hat{\mathcal W^-}}_L=\Sigma^{\hat{\phi}^+ \hat{\mathcal{W}}^-}=\Sigma^{\hat{\phi}^+\hat{\phi}^-}\\[16pt]
\Sigma^{\hat{\mathcal Z}\hat{\mathcal Z}}_L&=&c_{H\Box}c_{HWB}\frac{3g_1g_2v_T^6}{32\pi^2\epsilon}(g_1^2+g_2^2)\\[16pt]
\Sigma^{\hat{\mathcal Z}\hat \chi}&=&-c_{H\Box}c_{HWB}\frac{9ig_1g_2v_T^5}{64\pi^2\epsilon}\sqrt{g_1^2+g_2^2}\\[16pt]
\Sigma^{\hat\chi\hat\chi}&=&-c_{H\Box}c_{HWB}\frac{3g_1g_2v_T^4k^2}{16\pi^2\epsilon}\\[16pt]
T^H&=&c_{H\Box}c_{HWB}\frac{g_1g_2v_T^7}{128\pi^2\epsilon}(9g_1^2+9g_2^2+4\lambda\xi)
\end{array}
\end{equation}

\newpage
\subsection*{$c_{HD}c_{HB}$ contributions}
\begin{equation}
\begin{array}{rcl}
\Sigma^{\hat{\mathcal A}\hat{\mathcal A}}_L&=&0=\Sigma^{\hat{\mathcal A}\hat{\mathcal Z}}_L\\[16pt]
\Sigma^{\hat{\mathcal Z}\hat{\mathcal Z}}_L&=&c_{HD}c_{HB}\frac{g_1^2v_T^6}{128\pi^2\epsilon}[12\lambda+g_1^2(9+2\xi)+g_2^2(15+4\xi)]\\[16pt]
\Sigma^{\hat{\mathcal W^+}\hat{\mathcal W^-}}_L&=&c_{HD}c_{HB}\frac{3g_1^2g_2^2v_T^6}{128\pi^2\epsilon}\\[16pt]
\Sigma^{\hat{\mathcal Z}\hat \chi}&=&-c_{HD}c_{HB}\frac{3ig_1^2v_T^5}{512\pi^2\epsilon\sqrt{g_1^2+g_2^2}}[16\lambda+3g_1^2(5+\xi)+g_2^2(21+5\xi)]\\[16pt]
\Sigma^{\hat{\phi}^+ \hat{\mathcal{W}}^-}&=&-c_{HD}c_{HB}\frac{3g_1^2g_2v_T^5}{64\pi^2\epsilon}\\[16pt]
\Sigma^{\hat\chi\hat\chi}&=&c_{HD}c_{HB}\frac{g_1^2v_T^4}{256\pi^2\epsilon}[v_T^2(15g_1^2+15g_2^2-4\lambda\xi)-4(6+\xi)k^2]\\[16pt]
\Sigma^{\hat{\phi}^+\hat{\phi}^-}&=&c_{HD}c_{HB}\frac{3g_1^2v_T^4}{32\pi^2\epsilon}[(g_1^2+g_2^2)v_T^2-k^2]\\[16pt]
T^H&=&c_{HD}c_{HB}\frac{g_1^2v_T^7}{512\pi^2\epsilon}(39g_1^2+39g_2^2-4\lambda\xi)
\end{array}
\end{equation}

\vfill
\subsection*{$c_{HD}c_{HW}$ contributions}
\begin{equation}
\begin{array}{rcl}
\Sigma^{\hat{\mathcal A}\hat{\mathcal A}}_L&=&0=\Sigma^{\hat{\mathcal A}\hat{\mathcal Z}}_L\\[16pt]
\Sigma^{\hat{\mathcal Z}\hat{\mathcal Z}}_L&=&c_{HD}c_{HW}\frac{g_2^2v_T^4}{128\pi^2\epsilon}[12\lambda+3g_2^2(7+2\xi)+g_1^2(15+4\xi)]\\[16pt]
\Sigma^{\hat{\mathcal W^+}\hat{\mathcal W^-}}_L&=&c_{HD}c_{HW}\frac{3v_T^6g_2^2(g_1^2-2g_2^2)}{128\pi^2\epsilon}\\[16pt]
\Sigma^{\hat{\mathcal Z}\hat \chi}&=&-c_{HD}c_{HW}\frac{3ig_2^2v_T^5}{512\pi^2\epsilon\sqrt{g_1^2+g_2^2}}[16\lambda+g_1^2(27+7\xi)+g_2^2(33+9\xi)]\\[16pt]
\Sigma^{\hat{\phi}^+ \hat{\mathcal{W}}^-}&=&c_{HD}c_{HW}\frac{3g_2v_T^5(-g_1^2+3g_2^2)}{128\pi^2\epsilon}\\[16pt]
\Sigma^{\hat\chi\hat\chi}&=&-c_{HD}c_{HW}\frac{3g_2^2v_T^4}{256\pi^2\epsilon}[v_T^2(-5g_1^2+g_2^2+4\lambda\xi)+4(4+\xi)k^2]\\[16pt]
\Sigma^{\hat{\phi}^+\hat{\phi}^-}&=&c_{HD}c_{HW}\frac{3g_2^2v_T^4}{32\pi^2\epsilon}[(g_1^2+g_2^2)v_T^2+k^2]\\[16pt]
T^H&=&c_{HD}c_{HW}\frac{3g_2^2v_T^7}{512\pi^2\epsilon}[13g_1^2+7g_2^2-4\lambda\xi]
\end{array}
\end{equation}

\newpage
\subsection*{$c_{HD}c_{HWB}$ contributions}
\begin{equation}
\begin{array}{rcl}
\Sigma^{\hat{\mathcal A}\hat{\mathcal A}}_L&=&0=\Sigma^{\hat{\mathcal A}\hat{\mathcal Z}}_L\\[16pt]
\Sigma^{\hat{\mathcal Z}\hat{\mathcal Z}}_L&=&c_{HD}c_{HWB}\frac{g_1g_2v_T^6}{128\pi^2\epsilon}[12\lambda+g_1^2(9+2\xi)+g_2^2(15+4\xi)]\\[16pt]
\Sigma^{\hat{\mathcal W^+}\hat{\mathcal W^-}}_L&=&c_{HD}c_{HWB}\frac{3g_1g_2^3v_T^6}{128\pi^2\epsilon}\\[16pt]
\Sigma^{\hat{\mathcal Z}\hat \chi}&=&c_{HD}c_{HWB}\frac{3ig_1g_2v_T^5}{512\pi^2\epsilon\sqrt{g_1^2+g_2^2}}[16\lambda+3g_1^2(5+\xi)+g_2^2(21+5\xi)]\\[16pt]
\Sigma^{\hat{\phi}^+ \hat{\mathcal{W}}^-}&=&-c_{HD}c_{HWB}\frac{3g_1g_2^2v_T^5}{64\pi^2\epsilon}\\[16pt]
\Sigma^{\hat\chi\hat\chi}&=&c_{HD}c_{HWB}\frac{g_1g_2v_T^4}{256\pi^2\epsilon}[v_T^2(15g_1^2+15g_2^2-4\lambda\xi)-4(6+\xi)k^2]\\[16pt]
\Sigma^{\hat{\phi}^+\hat{\phi}^-}&=&c_{HD}c_{HWB}\frac{3g_1g_2v_T^4}{32\pi^2\epsilon}[(g_1^2+g_2^2)v_T^2-k^2]\\[16pt]
T^H&=&c_{HD}c_{HWB}\frac{g_1g_2v_T^7}{512\pi^2\epsilon}[39g_1^2+39g_2^2-4\lambda\xi]
\end{array}
\end{equation}
\vfill

\subsection*{$c_{HB}c_{HW}$ contributions}
\begin{equation}
\begin{array}{rcl}
\Sigma^{\hat{\mathcal A}\hat{\mathcal A}}_L&=&0=\Sigma^{\hat{\mathcal A}\hat{\mathcal Z}}_L\\[16pt]
\Sigma^{\hat{\mathcal Z}\hat{\mathcal Z}}_L&=&c_{HB}c_{HW}\frac{g_1^2g_2^2v_T^6}{16\pi^2\epsilon}(3+\xi)\\[16pt]
\Sigma^{\hat{\mathcal W^+}\hat{\mathcal W^-}}_L&=&c_{HB}c_{HW}\frac{g_1^2g_2^2v_T^6}{64\pi^2\epsilon}(3+\xi)\\[16pt]
\Sigma^{\hat{\mathcal Z}\hat \chi}&=&-c_{HB}c_{HW}\frac{ig_1^2g_2^2v_T^5}{128\pi^2\epsilon(g_1^2+g_2^2)^\frac{3}{2}}(7g_1^2+5g_2^2)(3+\xi)\\[16pt]
\Sigma^{\hat{\phi}^+ \hat{\mathcal{W}}^-}&=&-c_{HB}c_{HW}\frac{g_1^2g_2v_T^5}{64\pi^2\epsilon}(3+\xi)\\[16pt]
\Sigma^{\hat\chi\hat\chi}&=&c_{HB}c_{HW}\frac{3g_1^2g_2^2v_T^6}{16\pi^2\epsilon}\\[16pt]
\Sigma^{\hat{\phi}^+\hat{\phi}^-}&=&c_{HB}c_{HW}\frac{3g_1^2g_2^2v_T^6}{16\pi^2\epsilon}\\[16pt]
T^H&=&c_{HB}c_{HW}\frac{3g_1^2g_2^2v_T^7}{16\pi^2\epsilon}
\end{array}
\end{equation}

\newpage
\subsection*{$c_{HB}c_{HWB}$ contributions}
\begin{equation}\label{eq:cHBcHW}
\begin{array}{rcl}
\Sigma^{\hat{\mathcal A}\hat{\mathcal A}}_L&=&0=
\Sigma^{\hat{\mathcal A}\hat{\mathcal Z}}_L\\[16pt]
\Sigma^{\hat{\mathcal Z}\hat{\mathcal Z}}_L&=&c_{HB}c_{HWB}\frac{g_1g_2v_T^6}{16\pi^2\epsilon}(g_1^2+g_2^2)(3+\xi)\\[16pt]
\Sigma^{\hat{\mathcal W^+}\hat{\mathcal W^-}}_L&=&c_{HB}c_{HWB}\frac{g_1g_2^3v_T^6}{64\pi^2\epsilon}(3+\xi)\\[16pt]
\Sigma^{\hat{\mathcal Z}\hat \chi}&=&-c_{HB}c_{HWB}\frac{iv_T^5}{128\pi^2\epsilon(g_1^2+g_2^2)^\frac{3}{2}}(9g_1^5g_2+17g_1^3g_2^3+10g_1g_2^5)(3+\xi)\\[16pt]
\Sigma^{\hat{\phi}^+ \hat{\mathcal{W}}^-}&=&-c_{HB}c_{HWB}\frac{g_1g_2^2v_T^5}{32\pi^2\epsilon}(3+\xi)\\[16pt]
\Sigma^{\hat\chi\hat\chi}&=&c_{HB}c_{HWB}\frac{g_1g_2v_T^4}{16\pi^2\epsilon}[v_T^2(6g_1^2+3g_2^2+\lambda\xi)-(3+\xi)k^2]\\[16pt]
\Sigma^{\hat{\phi}^+\hat{\phi}^-}&=&c_{HB}c_{HWB}\frac{g_1g_2v_T^4}{16\pi^2\epsilon}[v_T^2(6g_1^2+3g_2^2+\lambda\xi)-(3+\xi)k^2]\\[16pt]
T^H&=&c_{HB}c_{HWB}\frac{g_1g_2v_T^7}{16\pi^2\epsilon}[6g_1^2+3g_2^2+\lambda\xi]
\end{array}
\end{equation}
\vfill

\subsection*{$c_{HW}c_{HWB}$ contributions}
\begin{equation}
\begin{array}{rcl}
\Sigma^{\hat{\mathcal A}\hat{\mathcal A}}_L&=&0=\Sigma^{\hat{\mathcal A}\hat{\mathcal Z}}_L\\[16pt]
\Sigma^{\hat{\mathcal Z}\hat{\mathcal Z}}_L&=&c_{HW}c_{HWB}\frac{g_1g_2v_T^6}{32\pi^2\epsilon}(g_1^2+4g_2^2)(3+\xi)\\[16pt]
\Sigma^{\hat{\mathcal W^+}\hat{\mathcal W^-}}_L&=&c_{HW}c_{HWB}\frac{g_1g_2^3v_T^6}{32\pi^2\epsilon}(3+\xi)\\[16pt]
\Sigma^{\hat{\mathcal Z}\hat \chi}&=&-c_{HW}c_{HWB}\frac{ig_1v_T^5}{128\pi^2\epsilon(g_1^2+g_2^2)^\frac{3}{2}}(6g_1^4g_2+23g_1^2g_2^3+15g_2^5)(3+\xi)\\[16pt]
\Sigma^{\hat{\phi}^+ \hat{\mathcal{W}}^-}&=&-c_{HW}c_{HWB}\frac{3g_1g_2^2v_T^5}{64\pi^2\epsilon}(3+\xi)\\[16pt]
\Sigma^{\hat\chi\hat\chi}&=&c_{HW}c_{HWB}\frac{g_1g_2v_T^4}{16\pi^2\epsilon}[v_T^2(3g_1^2+6g_2^2+\lambda\xi)-(3+\xi)k^2]\\[16pt]
\Sigma^{\hat{\phi}^+\hat{\phi}^-}&=&c_{HW}c_{HWB}\frac{g_1g_2v_T^4}{16\pi^2\epsilon}[v_T^2(3g_1^2+6g_2^2+\lambda\xi)-(3+\xi)k^2]\\[16pt]
T^H&=&c_{HW}c_{HWB}\frac{g_1g_2v_T^7}{16\pi^2\epsilon}(3g_1^2+6g_2^2+\lambda\xi)
\end{array}
\end{equation}

%
%
%
%
%
%
%
%
%
%
%
\newpage
\section{Speeding up the tools}\label{app:speedingup}
There are aspects of the design of \FR, \FA, and \FC which are not conducive to the SMEFT in the background field formalism. These are outlined below and workarounds are presented.

\begin{enumerate}[leftmargin=*]
\item \FR: 
\begin{itemize}[leftmargin=*]
\item \FR is not well equipped for the doubling of fields required by the BGFM. In particular the covariant derivatives must be defined either independent of the implicitly defined covariant derivative or the use of unphysical fields may be used. This work uses both methods. Unphysical fields for the gluon, W, and B, fields are defined so that the doubling of fields is implicit in the unphysical field. The covariant derivative of the four-component real $\phi$ field is defined explicitly as per Eq.~\ref{eq:Dphi4}.
\item The function \texttt{ExpandIndices} is very slow when dealing with higher dimensional operators. As such field multiplets are defined in this work so that the indices are explicitly expanded. This dramatically speeds up the expansion of the Lagrangian and derivation of the Feynman rules, it comes at the cost that families of fermions are no longer grouped together and instead are defined separately for the Feynarts level.
\item The increased multiplicity of fields due to the BGFM means deriving all the rules from the Lagrangian takes a very long time. This can be sped up with the internally defined parameter \texttt{epsf} ($\epsilon_f$). By making the substitution $\psi\to\epsilon_f \psi $, where $\psi$ is meant to represent all of the fields of the Lagrangian, and taking a Taylor series in $\epsilon_f$ we can select only three and four point functions, for example, for the derivation of the one loop Ward Identities contained in this and the accompanying paper. It is important to note that because the Maclaurin series is well defined is it significantly faster to define one's own Maclaurin series in terms of the derivative function of Mathematica than to use the series function. This is dramatically faster than the use of the \FR built in function \texttt{GetInteractionTerms} with \texttt{MaxParticles} and \texttt{MinParticles}, and has been cross checked against these functions. This is demonstrated in the ancillary files in the Mathematica notebook, \texttt{feynmanrules.nb}.
\end{itemize}
\item Feynarts/Formcalc:
\begin{itemize}[leftmargin=*]
\item Formcalc's \texttt{CalcFeynAmp} can be very slow when dealing with so many vertices and, more particularly, with vertices which have complicated coefficients for each Lorentz form. This can be evaded by reading the ``.mod'' file into Mathematica, copying all entries in \texttt{M\$CouplingMatrices}, which are generally sums of couplings/Wilson Coefficients, and replacing them with a single variable which are substituted back in after \texttt{CalcFeynAmp} has been run. For example, the vertex $\hat h^3$, is written in the Feynarts Mod file as:
\begin{equation}
3i gc_{60}[10c_H v^2 + \lambda(-4 - 12c_{H\Box}v^2 + 3c_{HD}v^2)]+2i(4c_{H\Box} - c_{HD})gc_{60}p_1\cdot p_2+\cdots
\end{equation}
where the $gc_i$ are internally defined substitutions from Feynrules. Instead we write this vertex as:
\begin{equation}
r_1+r_2\ p_1\cdot p_2+\cdots
\end{equation}
This is essentially taking the Feynrules $gc$ substitutions one step further. This has the effect of allowing Formcalc to simplify amplitudes that otherwise get hung up and never complete calculating. For example, without this step the calculation of $V\to V'$ with two internal vector lines in the general $R_\xi$ gauge never completes without this simplification of the .mod file. With the simplification it runs in only a few minutes. As color indices are not explicitly expanded, any rule involving color is skipped to simplify the process. This has worked well for the derivation of the Ward Identities, but may not be enough for more complex QCD processes. This is demonstrated in the ancillary files in the Mathematica notebook \texttt{rulesrearrange.nb} contained in the subdirectory \texttt{SMEFT6\_3and4pts}.
\item Formcalc's built in function \texttt{UVDivergentPart} fails to correctly produce the the UV divergent part of Passarino Veltman (PV) functions with too-high tensor rank, such as those that occur from the extra momentum insertions from higher dimensional operators in EFTs. Instead the Formcalc results may be written to file in terms of PV functions, read back into a cleared Mathematica terminal, and then the UV divergent part is derived using \textsc{Package X} \cite{Patel:2016fam}. Alternatively the PV functions can be reduced to scalar functions using the built in functions in Formcalc, but this was found to be too slow to be practical. This is performed at the end of the notebook, \texttt{ampprint.nb}, in the \texttt{SMEFT6\_3and4pts} subdirectory of the ancillary files.
\end{itemize}
\end{enumerate}
These steps are demonstrated in the ancillary files which include Mathematica notebooks which perform all of the steps mentioned above. The ancillary files are organized as follows:
\begin{enumerate}
\item The Feynrules package is contained in four .fr files, \texttt{BGSMEFT.fr}, \texttt{conventions.fr}, \texttt{fieldsandmetrics.fr}, and \texttt{lagrangians.fr}.
\item The notebook, \texttt{FeynmanRules.nb} can be used to derive the Feynman rules from the package, and includes the steps for quickly reducing the Lagrangian to include only $n$-point vertices. 
\item The subdirectory, \texttt{SMEFT6\_3and4pts} contains the \FA model files for 3 and 4 point tree level vertices derived from the package and including only dimension-six operators to $\mathcal O(1/\Lambda^2)$. These are \texttt{SMEFT6\_3and4pts.gen} and \texttt{SMEFT6\_3and4pts.mod}.
\item The Mathematica notebook \texttt{rulesrearrange.nb} simplifies the \FA model files as discussed above. It is contained in the subdirectory \texttt{SMEFT6\_3and4pts}.
\item The Mathematica notebook \texttt{ampprint} generates the two point functions and tadpole at one loop using \FA and \FC. The UV divergent parts of the amplitudes are determined using \textsc{Package-X} and printed on screen.
\end{enumerate}

\newpage
\bibliographystyle{JHEP}
\bibliography{ref}
\end{document}